\documentclass[reprint]{elsarticle}
\usepackage{graphicx}
\usepackage{amsmath}
\usepackage{amssymb}

\journal{Physics Letters A}

\begin{document}

\begin{frontmatter}

\title{Probabilistic Generation of Quantum Contextual Sets}

\author[label1]{Norman D.~Megill}


\address[label1]{Boston Information Group, 19 Locke Ln., Lexington,
MA 02420, USA}

\author[label2]{Kre{\v s}imir Fresl}


\address[label2]{Department of Mechanics, Faculty of Civil Engineering,
University of Zagreb,  Croatia}

\author[label3]{Mordecai Waegell}


\author[label3]{P.\ K.\ Aravind}


\address[label3]{Physics Department, Worcester Polytechnic Institute,
Worcester, MA 01609, U.~S.~A.}

\author[label4]{Mladen Pavi\v ci\'c}


\address[label4]{ITAMP at Physics Department at
Harvard University and Harvard-Smithsonian Center for
Astrophysics, Cambridge, MA 02138 USA \&\\
Chair of Physics, Faculty of Civil Engineering,
University of Zagreb, Croatia}

\date{\today}

\begin{abstract}
We give a method for exhaustive generation of a huge number of Kochen-Specker
contextual sets, based on the 600-cell, for possible experiments and
quantum gates. The method is complementary to
our previous parity proof generation of these sets, and it gives all
sets while the parity proof method gives only sets with an odd number 
of edges in their hypergraph representation. Thus we obtain 
35 new kinds of critical KS sets with an even number of edges.
Using a random sample of the sets generated with our method,
we give a statistical estimate of the number of sets 
that might be obtained in an eventual exhaustive enumeration.
\end{abstract}

\keyword{Kochen-Specker sets \sep MMP hypergraphs \sep 600-cell}

\PACS{03.65.-w \sep 03.65.Aa \sep 42.50.Dv \sep 03.67.-a}

}
\end{frontmatter}

\section{Introduction}
\label{sec:intro}

Quantum contextuality is the property of a quantum system that
a result of any of its measurements might depend on other
compatible measurements that might be carried out on the system.
The so-called Kochen-Specker (KS) sets provide constructive proofs
of quantum contextuality and therefore provide straightforward
blueprints for their experimental setups. KS sets are likely to find
applications in the field of quantum information, similar to ones
recently found for the Bell setups in implementing
entanglements.~\cite{heinbriegel04,cabello-moreno-09}
The most recent result of A.~Cabello \cite{cabello-10-arxiv},
according to which local contextuality can be used to reveal
quantum nonlocality, supports this conjecture. Also our
most recent results
\cite{bdm-ndm-mp-fresl-jmp-10,mp-7oa-arXiv,pavicic-rmp09} show that
KS sets play an important role in Hilbert space description
of complex setups.

A series of KS experiments have been carried out in the last
ten years. The most recent ones made use of quantum gates and
employed recently developed quantum information techniques
of handling, manipulating, and measuring of qubits by means of
quantum circuits of such gates.

The  experiments were proposed, designed, and carried out for
spin$-\frac{1}{2}\otimes\frac{1}{2}$ particles (correlated
photons or spatial and spin neutron degrees of freedom).
\cite{simon-zeil00,michler-zeil-00,cabell-01,ks-exp-03,h-rauch06,cabello-fillip-rauch-08,b-rauch-09,k-cabello-blatt-09-arXiv,amselem-cabello-09-arXiv,liu-09,moussa-09-arXiv}
The KS sets that were used in these experiments were
from $2\times 2=4$-dim Hilbert space. In particular they
were either from the 24-24 class of KS sets (set with 18 through 24
vectors and 9  through 24 orthogonal vector tetrads) or the
Mermin set.\ \cite{pmm-2-09}

They used specified vectors (e.g., \cite{peres})  and relied on
particular orientation of measurement devices along those vectors.
That limited possible implementations of a given KS set.
Therefore in~\cite{pmmm04c,pmm-2-09} we exhaustively generated
all KS sets from the 24-24 class without ascribing coordinates to
Hilbert vector (states, wave function) components. That was done by
means of McKay-Megill-Pavicic (MMP) hypergraph representation
(MMP diagrams). For these hypergraphs it is only important that the
equations that determine vector components of a setup have solutions.
Solutions  themselves can be determined by an algorithm, and
observables need not be grouped or particularly chosen.
E.g., in both 3-dim (spin-1, qutrits) and 4-dim (spin-$\frac{3}{2}$)
KS setups we can make use of generalized Stern-Gerlach devices
\cite{anti-shimony} with outputs corresponding to vector components.

Most recently
\cite{mp-nm-pka-mw-11,waeg-aravind-megill-pavicic-11}
we generated millions of KS sets from a 4-dim 60-75 KS set we obtained
from the so-called  600-cell (the 4-dimensional analog
of the icosahedron)\cite{aravind10-arxiv}. Since they
all stem from this single 60-75 set and since no set from the 24-24
class belongs to it, we call it the {\em 60-75 KS class}. The
experimental implementations of the sets
belonging to this class are straightforward although demanding.
For instance, we let a spin-$\frac{3}{2}$ systems through a series
generalized Stern-Gerlach devices, enabling control over
outcoming directions of particles.~\cite{anti-shimony}
The approach can also be used to make quantum gates that must
be purely quantum for whatever state it applies to.

For any experimental application it is not viable to consider all
possible millions of sets but only those that can be experimentally
distinguished. Hence, we extract critical non-redundant
non-isomorphic KS sets with 26 to 60 vectors from all possible
60-75 KS sets.  ``Critical'' means that they are minimal in the sense
that no orthogonal tetrads can be removed without causing the KS
contradiction to disappear. We found several thousand critical KS
setups that have no experimental redundancies.

In \cite{waeg-aravind-megill-pavicic-11} we developed a
method of exhaustive generation of all those KS sets that allow
the so-called parity proofs (see below). However, the parity proofs
are applicable only to the sets with an odd number of tetrads
of orthogonal vectors, and the aforementioned generation
gives only such sets.

In this paper, we describe a method for
generating all KS sets from the 60-75 KS class,
in particular those sets that we cannot obtain
by our parity-proof-generation method.
While in principle the method is exhaustive, full generation
is at present too demanding for even a large supercomputing
cluster.  Instead, we used random samples of the search
space and applied Bernoulli trial probability analysis to
obtain expected means and confidence intervals.
We obtained these samples using techniques of graph theory and quantum
mechanical lattice theory. Also, since the parity proof method
is faster for obtaining critical KS sets with odd number of
vector tetrads (edges in hypergraphs), we concentrate on even
numbers of tetrads (see Table \ref{table:even}). In this sense our
probabilistic generation method and our parity-proof-generation
method are complementary. Also our present probabilistic generation
method is better at finding a large number of critical sets of
both kinds since it does not depend on the values ascribed to
vectors (vertices in hypergraphs) from the sets.

This study extends our preliminary results reported in
Ref.~\cite{mp-nm-pka-mw-11}.  We will describe the improved algorithms
that have allowed us to go beyond the results of that study and survey a
huge number of possible tetrads from 1 through 75.  In the random sample
used for our survey, they ranged from 26-13 (and suspected to be the
smallest) to a very large one, 60-41.  In addition, based on
statistical extrapolation from our sample, we give an estimation method 
according to which there might be a practically unlimited number 
($\approx 4.3\cdot 10^{12}$, Fig.~\ref{fig:stat}) of non-isomorphic 
critical KS sets that are subsets of the 60-75 set. The 
method is however esentially classical, so it might
happen that a future exahustive generation will give far 
less numbers of critical sets. If it does, that will show 
to which extent quantum data differ from classical 
estimations. If it confirms our estimation then we will 
have a powerful tool for estimating the reliability of 
random generation of critical KS sets. For non-critical 
KS sets the exahustive generation of sets with 63 to 75 
tetrads already confirmed our estimation; we give 
comparative numerical values in Sec.~\ref{sec:con}. 

Finally, we will summarize the overall picture of the critical KS sets we
found, describe patterns we have observed in their relative distribution
versus number of vectors and tetrads, and list some open questions about
whether others that we haven't found yet exist and whether, for some
sizes, we have exhausted all possible isomorphism classes.

We make use of theory and algorithms from several
disciplines: quantum mechanics, lattice theory, graph theory, and
geometry.  Thus in the context of our study,
the  term ``vertex'' is synonymous
with the terms ``ray,'' ``atom,'' ``1-dim subspace,'' and ``vector''
that appear in the literature; ``edge''  with the terms
``base,''  ``block,'' and ``tetrad (of mutually orthogonal vectors);'' and
``MMP hypergraph'' with the terms ``MMP diagram'' and
``KS sets.''

\section{Results}
\label{sec:res}

The {\em Kochen-Specker\/  {\rm (KS)} theorem} states that a
quantum system cannot in general possess a definite value of a
measurable property prior to measurement, and quantum
measurements (essentially detector clicks) carried out on
quantum systems cannot always be ascribed predetermined
values (say 0 and 1). This means that two measurements of
the same observable of the same system sometimes
must yield different outcomes in different contexts.
This is called the {\em quantum contextuality\/}.
One way of proving the theorem is to
prove the existence of KS sets, i.e., to provide
algorithms for their constructive generation. The more
abundant they are, the more important the contextuality
of quantum mechanics appears to be.

{\em Kochen-Specker\/ {\rm (KS)} set\/} is a set of vectors $
|\psi_i\rangle,\psi_i'\rangle,\dots$ in ${\cal H}^n$, $n\ge 3$ to
which it is impossible to assign $1$'s and $0$'s in such a way that:
\begin{enumerate}
\item No two orthogonal vectors are both
assigned the value $1$;
\item In any subset of $n$ mutually orthogonal vectors, not all
them are assigned the value $0$.
\end{enumerate}

KS subsets of mutually orthogonal vectors in a 3-dim space we call
triads, in a 4-dim space tetrads, etc. KS set is a union of such
triads, tetrads, etc. of vectors.  They can be represented by means of
MMP hypergraph defined below. In a KS  set, the vectors correspond
to vertices and the tetrads to edges of MMP hypergraphs.

We define MMP hypergraphs as follows\ \cite{pmmm04c}
\begin{itemize}
\item[(i)] Every vertex belongs to at least one edge;
\item[(ii)] Every edge contains at least 3 vertices;
\item[(iii)] Edges that intersect each other in $n-2$
         vertices contain at least $n$ vertices.
\end{itemize}

This definition enables us to formulate algorithms for exhaustive
generation of MMP hypergraphs.  We
work with subsets of the starting hypergraph, the 60-75 one, so
the job of generating the hypergraphs amounts to a creation of
all possible subsets of the 60-75 set with a specified number of edges
deleted.  The ``only'' difficulty we face is the shear size of these
generated subsets---we are dealing with a haystack of $2^{75}$ or 38
sextillion subsets, in which we wish to find certain ``needles'' i.e.
critical KS sets.  Our primary purpose is to survey the $2^{75}$
subsets to gain an overview of what critical sets are inside.

The hypergraphs we obtain reflect only the orthogonal structure
of KS sets and do not in any way refer to the vector components of the
original 60-75 KS set. This is yet another aspect in which the present
method differs from the parity-proof method we used in
\cite{waeg-aravind-megill-pavicic-11}, which relies on
the vector components of the vector in each KS sets that were
inherited from the original 60-75 set. For each hypergraph
we can however find appropriate vector components by our
program {\tt vectorfind} or by interval analysis we developed in
~\cite{pmmm04c}. These components need not have the
values the vector components have in the 60-75 set.

\font\1=cmss8
\font\2=cmssdc8
\font\3=cmr8
\font\4=cmss7

We encode MMP hypergraphs by means of alphanumeric
and other printable ASCII characters. Each vertex
is represented by one of the following
characters: {\11 2 3 4 5 6 7 8 9 A B C D E
F G H I J K L M N O P Q R S T U V W X Y Z a b c d
e f g h i j k l m n o p q r s t u v w x y z ! " \#
{\scriptsize\$} \% \& ' ( ) * - / : ; $<$ = $>$ ?
@ [ {\scriptsize$\backslash$} ] \^{} \_
{\scriptsize$\grave{}$} {\scriptsize\{} {\scriptsize$|$}
{\scriptsize\}} {\scriptsize\~{}}}\ , and then again all
these characters prefixed by `+', then prefixed by `++',
etc. There is no upper limit on the number of characters.

Each edge is represented by a string of characters that
represent vertices (without spaces). Edges are separated by
commas (without spaces). All edges in a line form a
representation of a hypergraph.  The order of the edges is
irrelevant---however, we shall often present them
starting with edges forming the biggest loop to facilitate
their possible drawing.  The line must end with a full stop.
Skipping of characters is allowed.

\begin{figure}[hbt]
\includegraphics[width=0.4\textwidth]{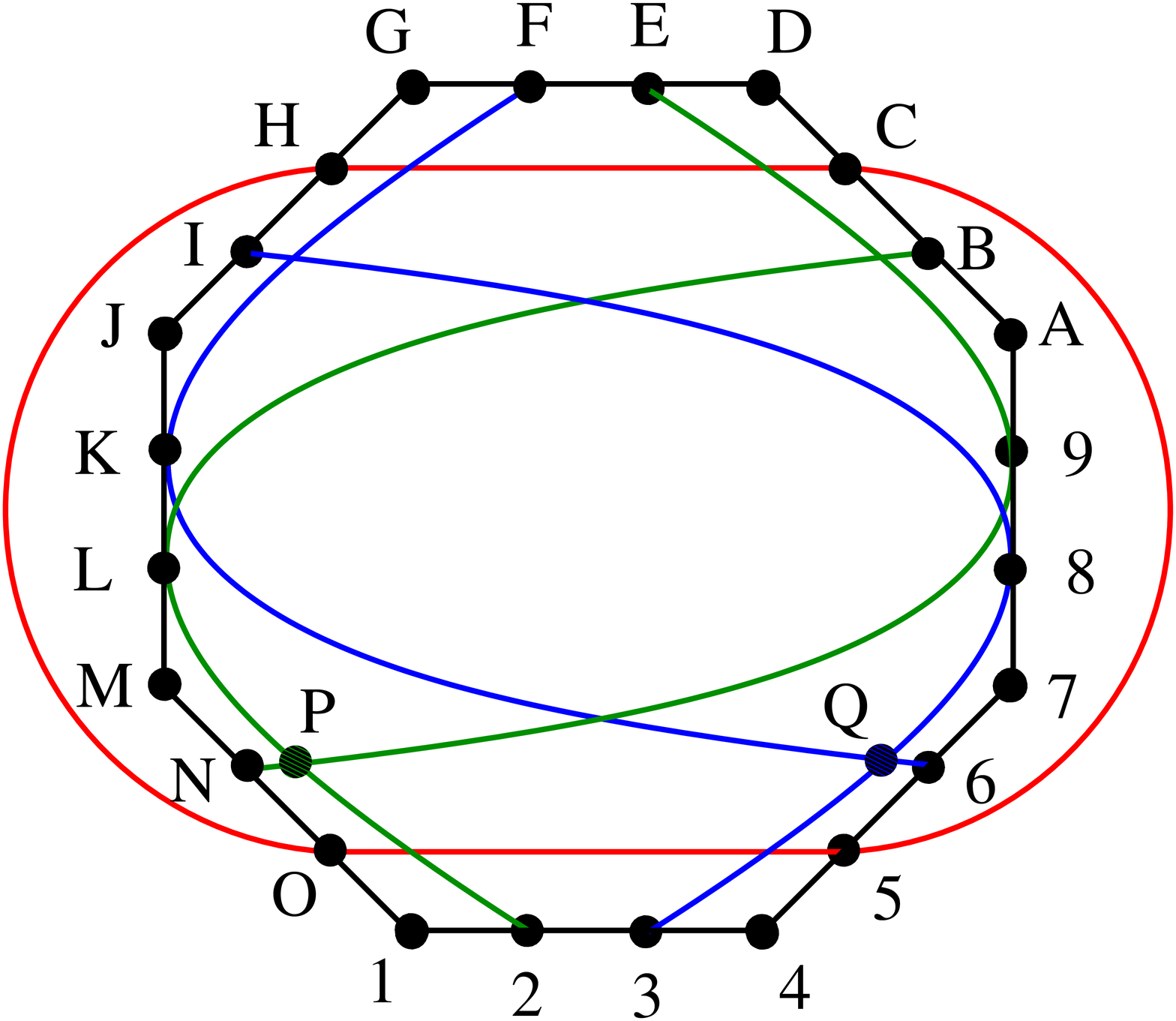}\qquad\qquad
\includegraphics[width=0.4\textwidth]{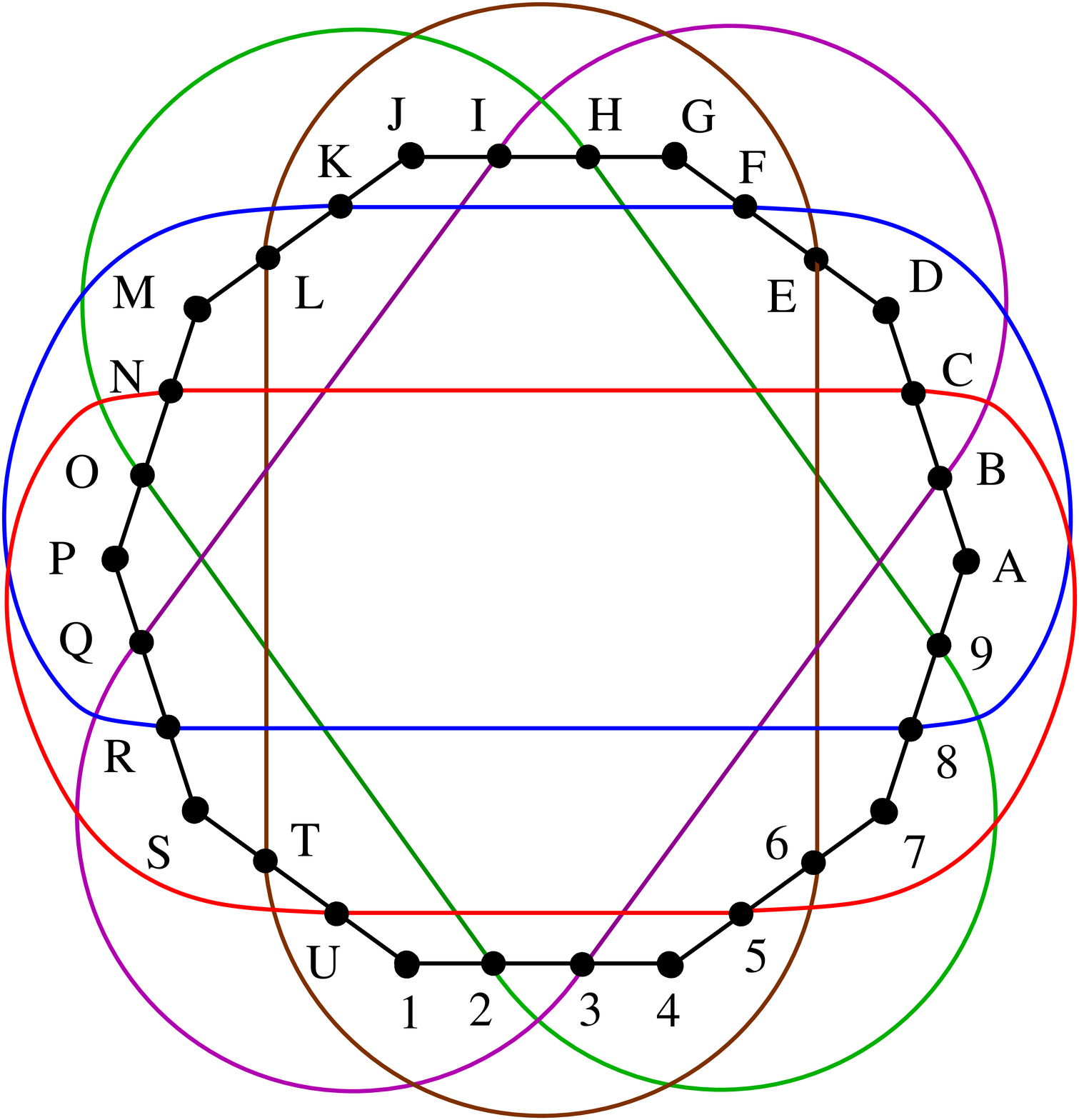}\quad%
\caption{\label{fig:30-15a}Critical KS sets 26-13 and 30-15.
They, as well as other figures with odd number of edges below,
vividly illustrate the parity proof: all vertices share two edges.}
\end{figure}

\begin{figure}[hbt]
\includegraphics[width=0.44\textwidth]{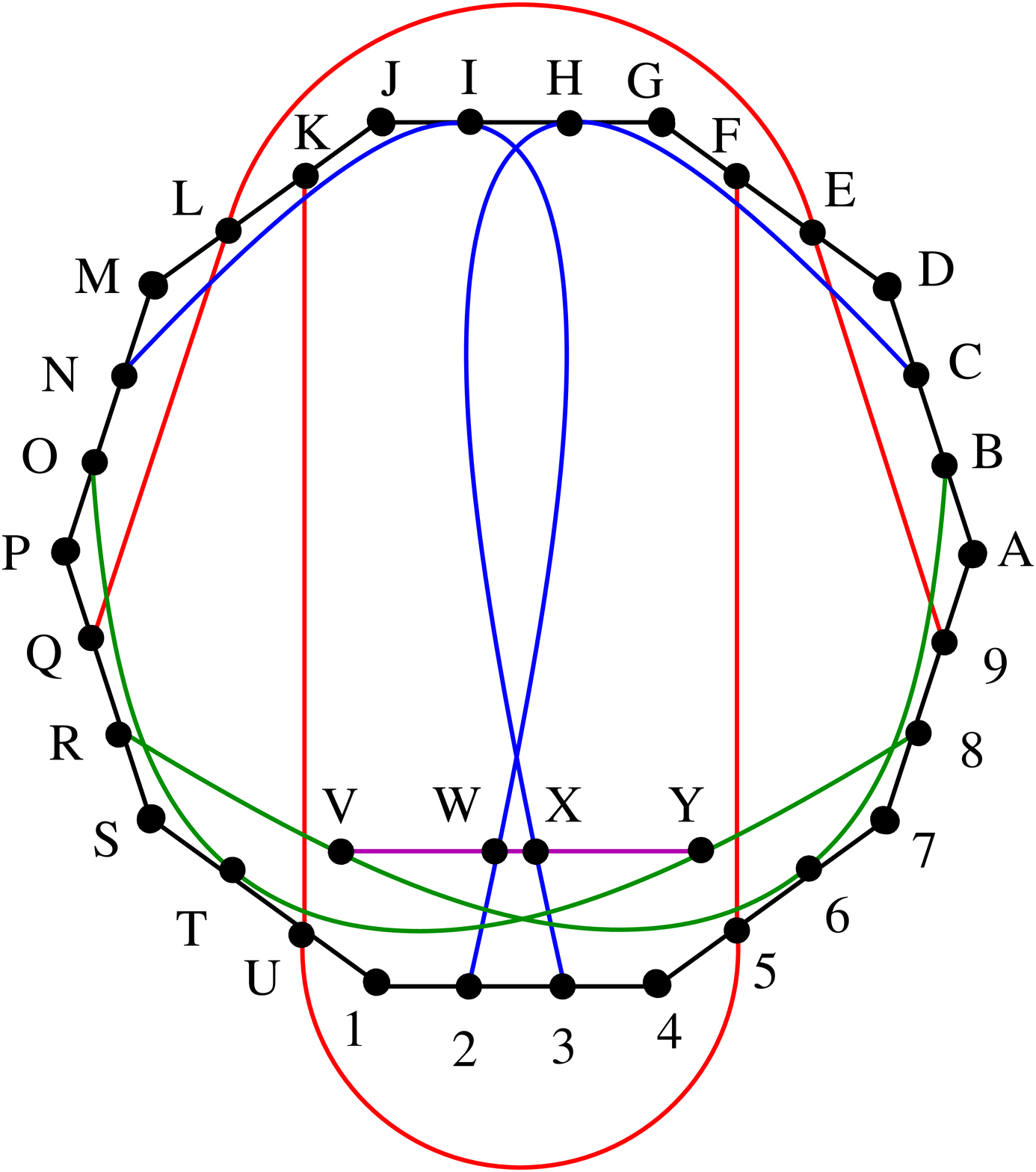}\qquad
\includegraphics[width=0.48\textwidth]{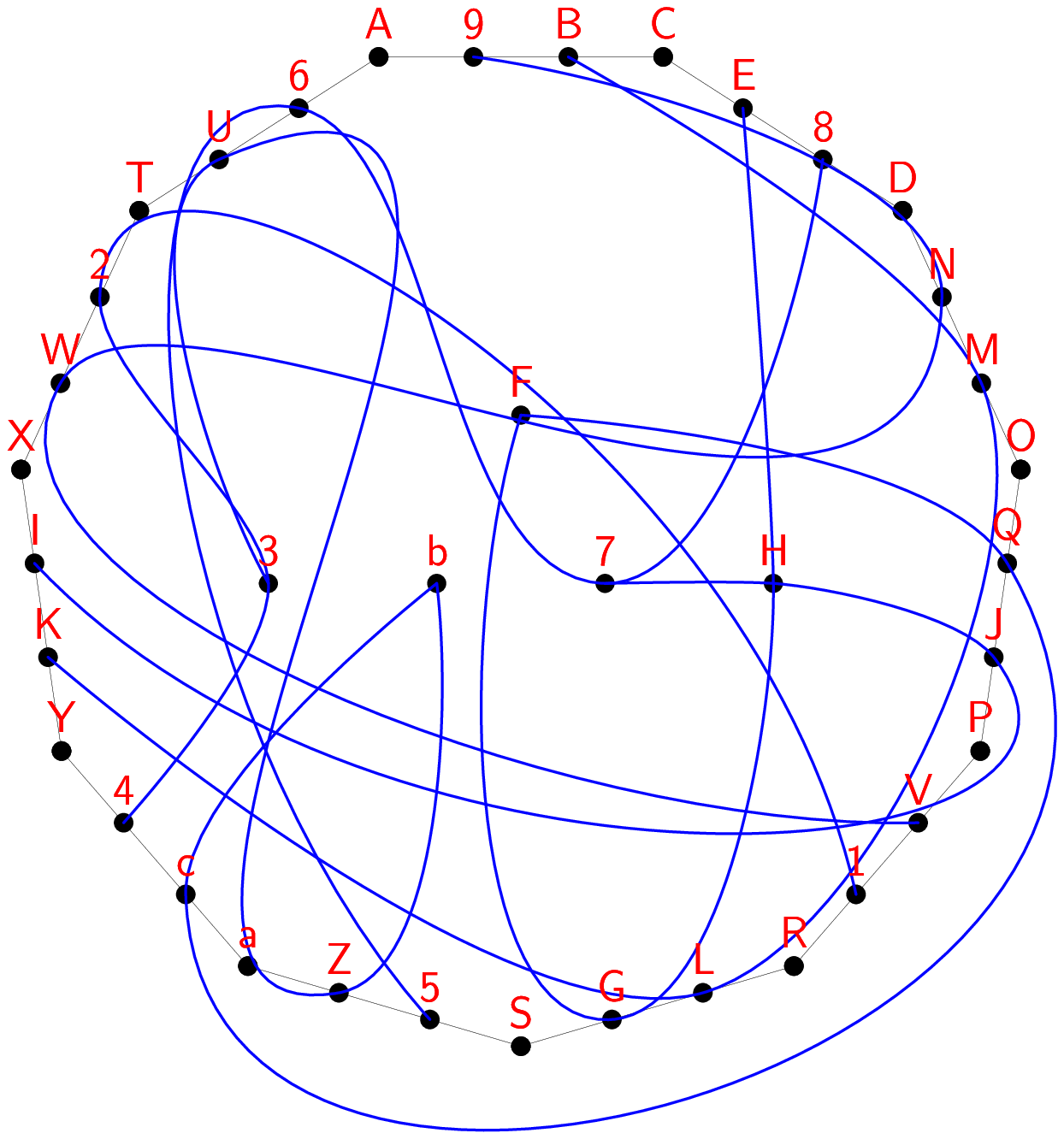}%
\caption{\label{fig:30-15b}Critical KS sets 34-17
and 38-19. The latter one is drawn by {\tt Asymptote} (vector graphic
language). By changing parameters one can interactively control and
change the shape of each edge line and unambiguously discern vertices
that share an edge.}
\end{figure}

In Figs.~\ref{fig:30-15a} and \ref{fig:30-15b} we show a
graphical representation of 3 critical KS hypergraphs from
\cite{mp-nm-pka-mw-11,waeg-aravind-megill-pavicic-11}
which are drawn by hand and a new one that is drawn by our
programs for automated drawing of MMP hypergraphs.
The MMP notation for  hypergraphs  in
Figs.~\ref{fig:30-15a} and \ref{fig:30-15b}$\,$(a)  is given in
\cite{mp-nm-pka-mw-11} and for 38-19 in Fig.~\ref{fig:30-15b}
it reads {\noindent\1  
A9BC,CE8D,DNMO,OQJP,PV1R,RLGS,SZ5a,ac4Y,YKIX,XW2T,TU6A,1234,5678,FGHE,IJH7,\break 
KLMB,VWN9,bcQF,bZU3.}

{\it Parity Proof.\/} By looking at the KS hypergraphs shown
in Figs.~\ref{fig:30-15a} and \ref{fig:30-15b}, we see
that we cannot ascribe values 0 and 1 to all vertices so
that in each edge we ascribe 1 to one of the vertices and
0 to the others. Namely, in all these hypergraphs,
each vertex shares exactly two edges, so there should be an
even number of 1s. At the same time, each edge must contain
one 1 by definition, and since there are an odd number of edges,
there should be an odd number of 1s---a contradiction. This is
the simplest form of the parity proofs of the Kochen-Specker
theorem on KS sets.

In these parity proofs, as well as in more involved ones we give
in \cite{waeg-aravind-megill-pavicic-11}, the essential
part of the proof is that we have an odd number of edges.
We found 90 kinds of such critical KS sets starting with a 26-13
and ending with a 60-41 one and presented them
qualitatively in Table 3 of
Ref.~\cite{waeg-aravind-megill-pavicic-11}.
These KS set we found by means of the parity proofs in
Ref.~\cite{waeg-aravind-megill-pavicic-11}, but not by
the algorithms we used for this paper, we indicated by
``$\otimes$'' in Table~\ref{table:even} below.

In Table~\ref{table:even} and Fig.~\ref{fig:stat}, we show the
distribution and numbers for each kind that were found by the
random sampling of our present survey.

We prove the impossibility of assigning 0 and 1 to vertices by means
our program {\tt states01}, which exhaustively tests
each hypergraph. For KS sets with an odd number of edges, we can
also quickly verify their KS property with the help of the parity
proof.

In the sample used for our survey, on average $3\cdot 10^8$ for
each of 13--63 edges (1--12 and 64--75 were exhaustively scanned), we
obtained 35 kinds of  critical KS sets with an even number of
edges, which cannot be obtained by the parity-proof method.
We also obtained 62 kinds of critical KS sets with an odd number of
edges. They are all among 90 kinds of KS critical sets we obtained by
the parity-proof method in \cite{waeg-aravind-megill-pavicic-11}.
Those that we did not obtain in our samples are indicated
by ``$\otimes$'' in Table~\ref{table:even}. Our scanning in
\cite{waeg-aravind-megill-pavicic-11} was designed to obtain as
many different kinds of critical sets as possible. So, we always stopped
scanning as soon as we found a new kind and therefore we cannot
estimate numbers of critical sets of each kind that would correspond
to to numbers obtained in this paper and shown in Table~\ref{table:even}.

\begin{table}[hbt]
\addtolength{\tabcolsep}{-4.25pt}
{\1\begin{tabular}{|c||c|c|c|c|c|c|c|c|c|c|c|c|c|c|c||c|c|c|c|c|c|c|c|c|c|} 
\hline&\multicolumn{15}{|c||}{\2critical KS sets with odd
  number of edges}&\multicolumn{9}{|c|}{\2\dots\ with even number of edges}\\
\hline \hline
{\2ver}&13&15&17&19&21&23&25&27&29&31&33&35&37&39&41&24&26&28&30&32&34&36&38&40\\[-2.5pt]
\hline \hline
26&1&&&&&&&&&&&&&&&&&&&&&&&\\[-2.5pt]
\hline\hline
30&&3&&&&&&&&&&&&&&&&&&&&&&\\[-2.5pt]
\hline
32&&&1&&&&&&&&&&&&&&&&&&&&&\\[-2.5pt]
\hline
33&&&2&&&&&&&&&&&&&&&&&&&&&\\[-2.5pt]
\hline
34&&&5&&&&&&&&&&&&&&&&&&&&&\\[-2.5pt]
\hline\hline
36&&&&11&&&&&&&&&&&&&&&&&&&&\\[-2.5pt]
\hline
37&&&&9&&&&&&&&&&&&&&&&&&&&\\[-2.5pt]
\hline
38&&&&6&10&&&&&&&&&&&&&&&&&&&\\[-2.5pt]
\hline
39&&&&&30&&&&&&&&&&&&&&&&&&&\\[-2.5pt]
\hline
40&&&&&38&10&&&&&&&&&&&&&&&&&&\\[-2.5pt]
\hline
41&&&&&22&5&&&&&&&&&&&&&&&&&&\\[-2.5pt]
\hline
42&&&&&6&16&3&&&&&&&&&1&&&&&&&&\\[-2.5pt]
\hline
43&&&&&&22&38&&&&&&&&&&&&&&&&&\\[-2.5pt]
\hline
44&&&&&&14&16&$\otimes$&&&&&&&&&&&&&&&&\\[-2.5pt]
\hline
45&&&&&&3&5&32&&&&&&&&&2&&&&&&&\\[-2.5pt]
\hline
46&&&&&&1&3&130&$\otimes$&&&&&&&&&3&&&&&&\\[-2.5pt]
\hline
47&&&&&&&1&74&$\otimes$&&&&&&&&&6&&&&&&\\[-2.5pt]
\hline
48&&&&&&&2&19&9&$\otimes$&&&&&&&&11&&&&&&\\[-2.5pt]
\hline
49&&&&&&&$\otimes$&$\otimes$&11&1&&&&&&&&3&9&&&&&\\[-2.5pt]
\hline
50&&&&&&&1&$\otimes$&7&13&&&&&&&&&39&&&&&\\[-2.5pt]
\hline
51&&&&&&&&$\otimes$&1&33&$\otimes$&&&&&&&&19&18&&&&\\[-2.5pt]
\hline
52&&&&&&&&$\otimes$&$\otimes$&37&33&&&&&&&&4&69&&&&\\[-2.5pt]
\hline
53&&&&&&&&$\otimes$&$\otimes$&11&114&$\otimes$&&&&&&&1&73&45&&&\\[-2.5pt]
\hline
54&&&&&&&&$\otimes$&$\otimes$&$\otimes$&153&16&&&&&&&&26&275&&&\\[-2.5pt]
\hline
55&&&&&&&&&$\otimes$&1&56&158&$\otimes$&&&&&&&5&339&25&&\\[-2.5pt]
\hline
56&&&&&&&&&$\otimes$&$\otimes$&21&241&28&&&&&&&&136&262&&\\[-2.5pt]
\hline
57&&&&&&&&&&$\otimes$&1&133&378&&&&&&&&54&448&45&\\[-2.5pt]
\hline
58&&&&&&&&&&$\otimes$&$\otimes$&30&678&27&&&&&&&2&256&493&\\[-2.5pt]
\hline
59&&&&&&&&&&&$\otimes$&2&308&381&&&&&&&1&55&864&16\\[-2.5pt]
\hline
60&&&&&&&&&&&$\otimes$&$\otimes$&48&562&1&&&&&&&5&316&145\\
\hline
\end{tabular}}
\caption{List of KS critical sets we obtained in this paper. By $\otimes$
we indicate the existence of KS critical sets (at least
one set) we obtained in \cite{waeg-aravind-megill-pavicic-11} by
the parity proof method. The average sample sizes of
$3\cdot 10^8$ sets used here were too small
to obtain them by our algorithms/programs.}
\label{table:even}
\end{table}

\begin{figure}[hbt]
\includegraphics[width=0.45\textwidth]{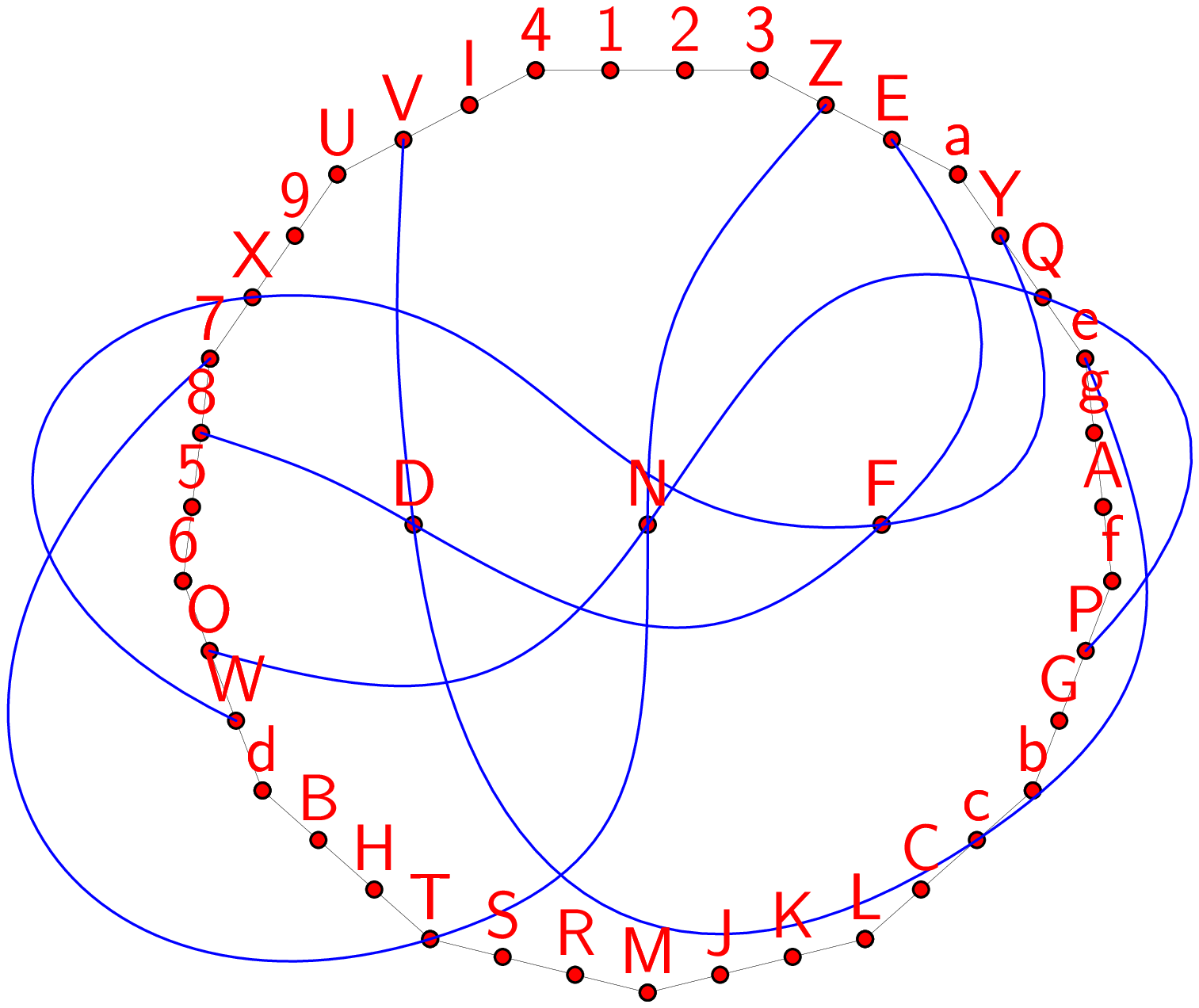}\qquad \qquad
\includegraphics[width=0.37\textwidth]{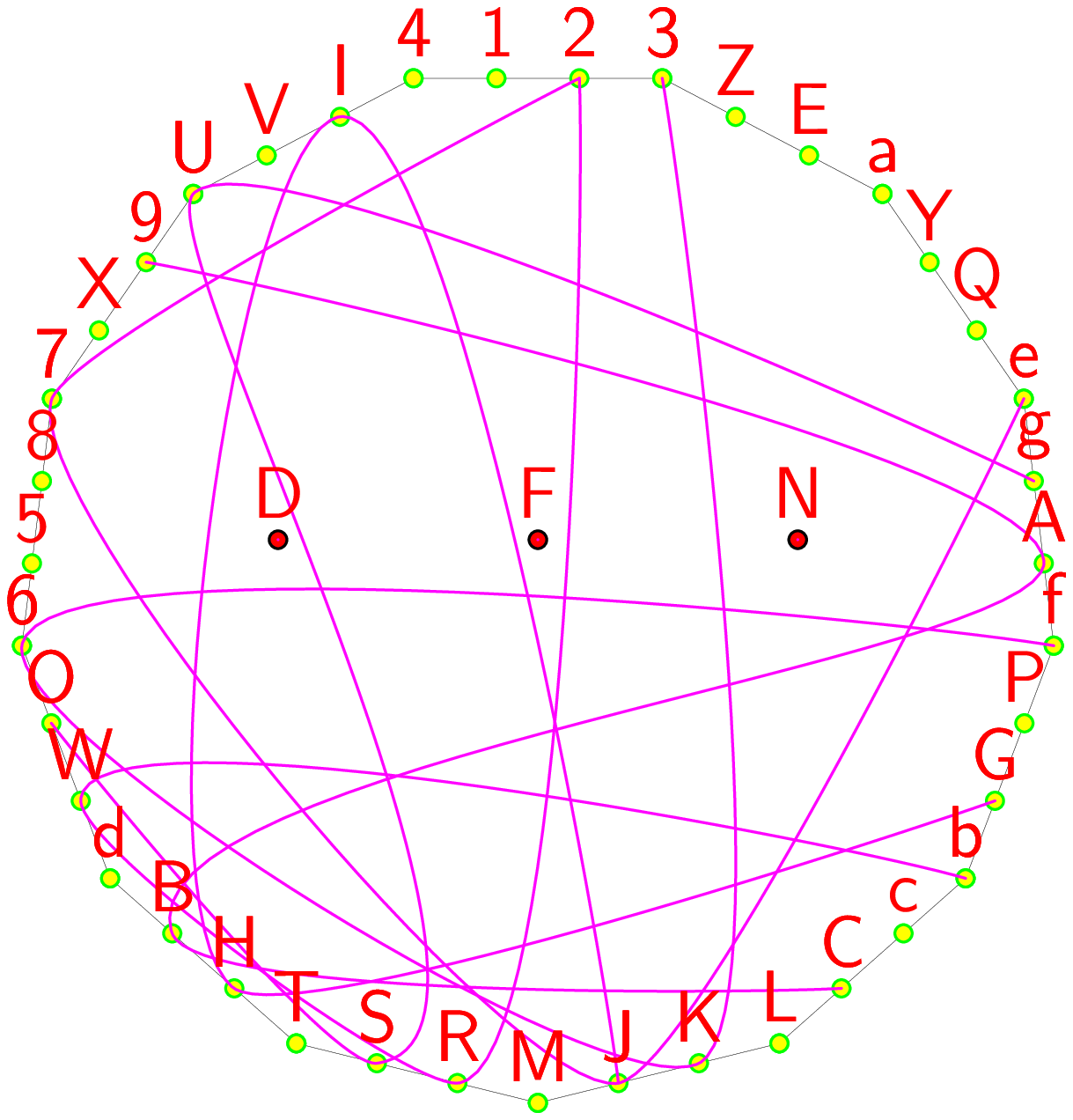}
\caption{\label{fig:even24-30a} Critical KS sets with even number of
edges. (a,b) presentations of 42-24 using the
biggest loop (13-gon): (a) 13-gon with only free edges; (b) with only
span edges; tension of the lines in (b) is higher for better
discerning of the edges.}
\end{figure}

\begin{figure}[hbt]
\includegraphics[width=0.45\textwidth]{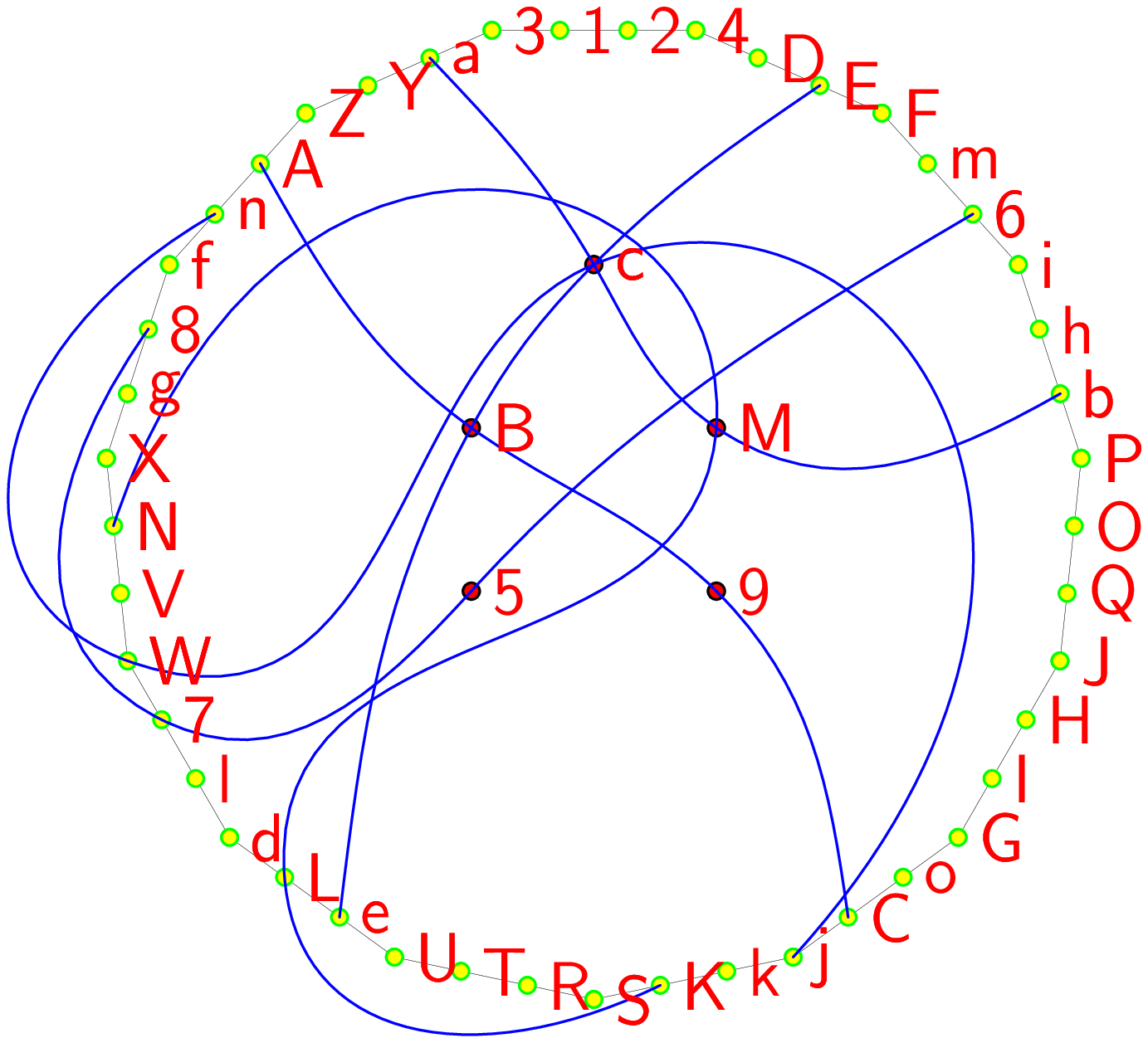}\qquad
\includegraphics[width=0.42\textwidth]{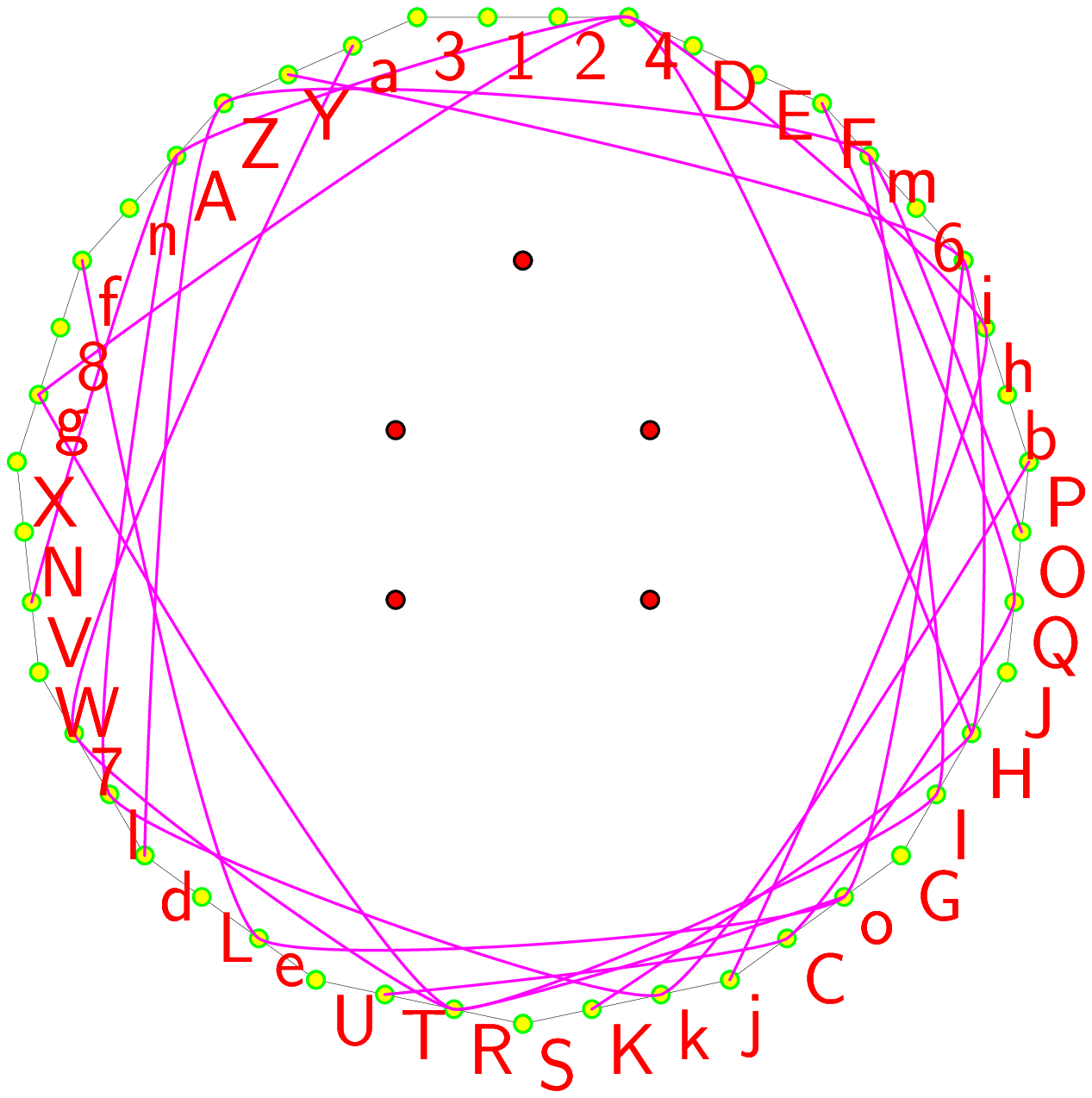}
\caption{\label{fig:even24-30b} (a,b) 50-30 (15-gon): (c) free edges only;
(d) span edges only. We can add or delete vertices and edges [as in
of Fig.~\ref{fig:even24-30a} and Fig.~\ref{fig:54-34b}]
according to an algorithm so as to obtain sub-hypergraphs for
various applications, e.g, finding Hilbert space equations for quantum
systems.~\cite{bdm-ndm-mp-fresl-jmp-10,mp-7oa-arXiv}}
\end{figure}

\begin{figure}[hbt]
\begin{center}
\includegraphics[width=0.99\textwidth]{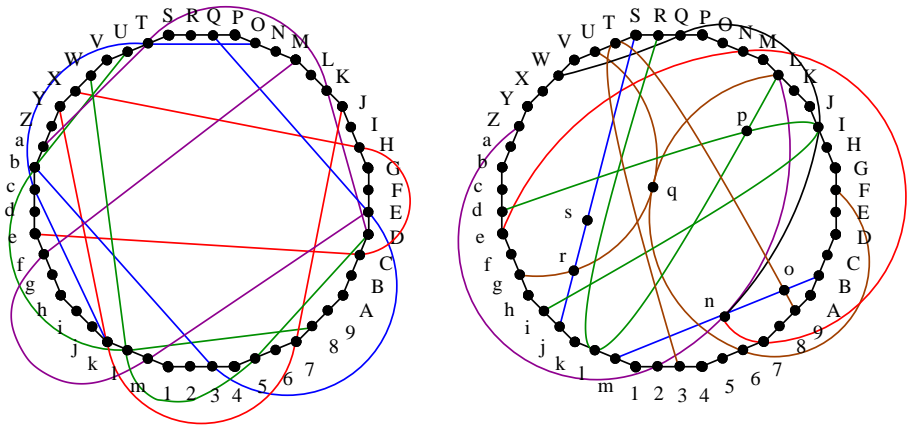}
\end{center}
\caption{\label{fig:54-34a}  A critical 54-34 KS set shown in two
figures. (a) A maximal 16-gon + span edges; (b) 16-gon + free edges.}
\end{figure}

\begin{figure}[hbt]
\begin{center}
\includegraphics[width=0.42\textwidth]{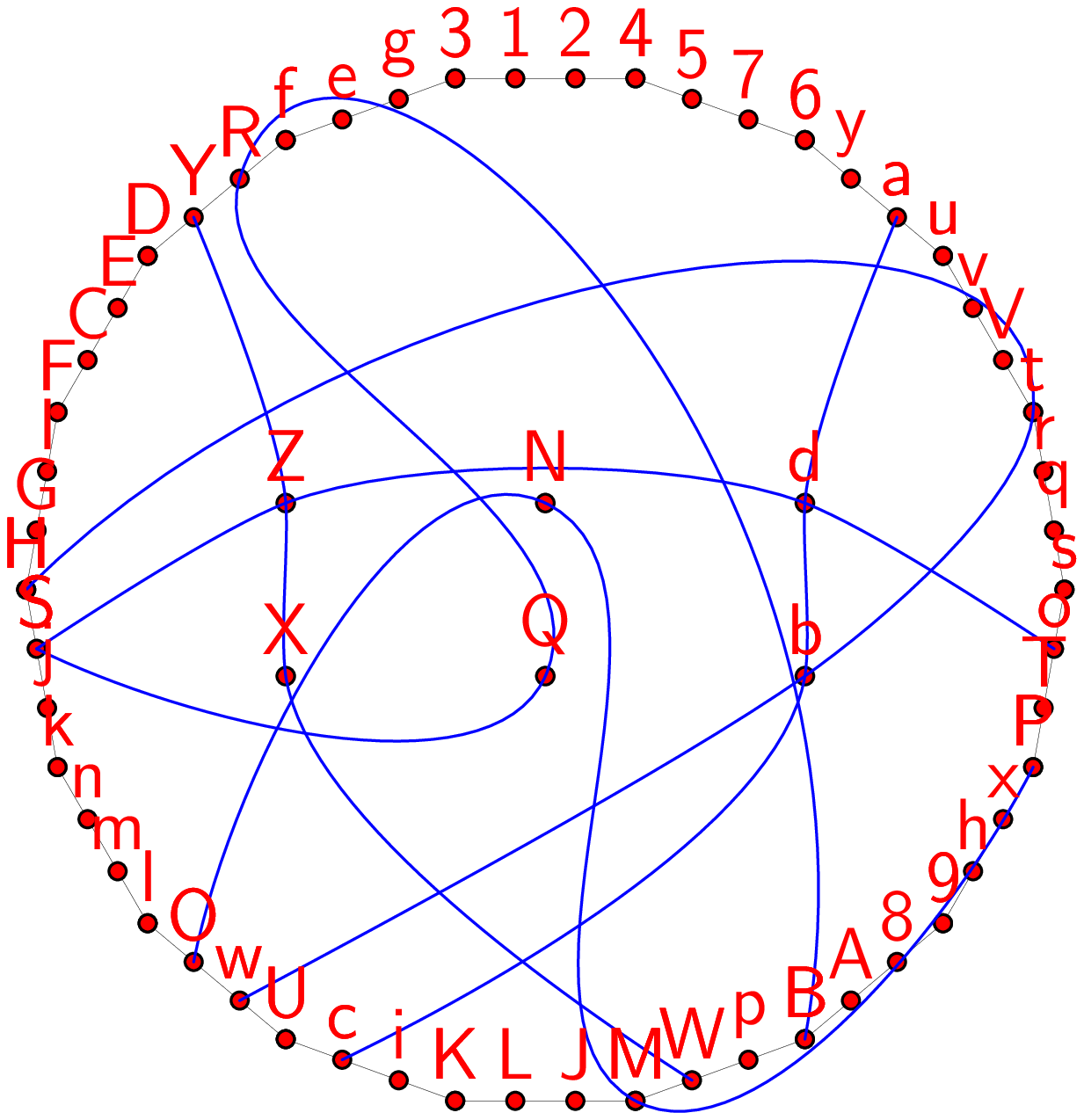}\qquad
\includegraphics[width=0.48\textwidth]{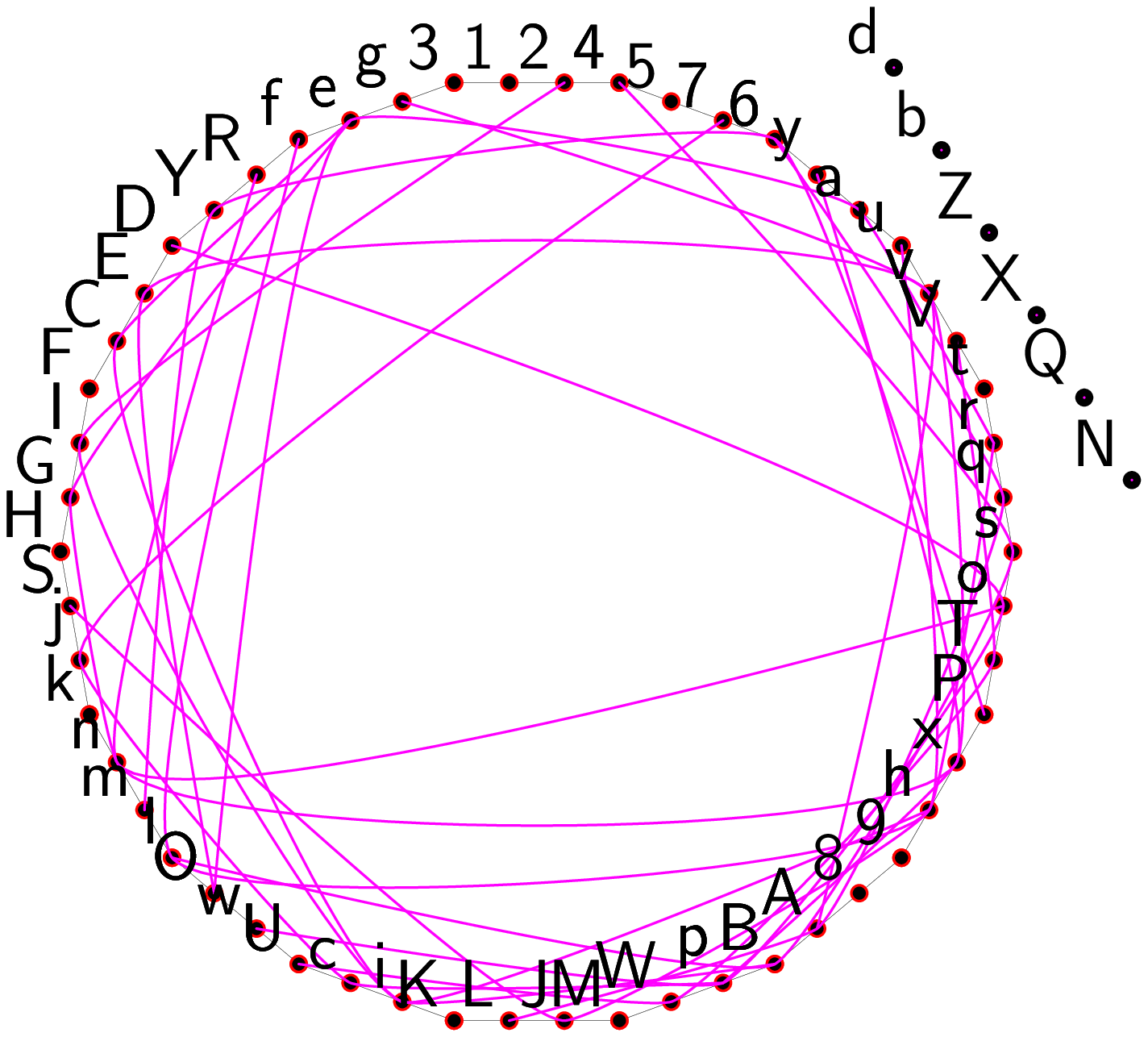}
\end{center}
\caption{\label{fig:54-34b} A critical 60-40 set;
(a) A maximal 18-gon + free edges; (b) 18-gon + span edges. Letters
and edges might appear overcrowded, but the MMP hypergraph
offers a clear alternate representation for each of them. Also, by means of our
programs, they can be easily discerned textually and visually.
Actually, when an MMP figure of a KS set is shown as a figure, we
need not give ASCII characters at all---we can always ascribe them
to vertices in an arbitrarily chosen way later on.}
\end{figure}

In Figs.~\ref{fig:even24-30a}, \ref{fig:even24-30b},
\ref{fig:54-34a},  and \ref{fig:54-34b}, we show some
further examples of hypergraphs automatically drawn
by our program written in {\tt Asymptote}.

In the MMP notation for these (shown below), the
numbers in parentheses [$(n)$] are the size of
the maximal loops ($n$-gons) that the
edges can form. The first $n$ edges (tetrads) form $n$-gons.
We obtain them by our program {\tt loopbig\/}. Additional examples of
hypergraphs of each kind not given here are listed in \ref{sec:app}.

There are three types of edges in an MMP hypergraph
\newcommand{\compactlist}{\setlength{\itemsep}{0pt} \setlength{\parskip}{0pt} \setlength{\leftskip}{7.7em}}
\begin{itemize}
\compactlist
\item[\hbox to 80pt{\em Polygon edges\hfill}]those that form the
  $n$-gon.
\item[\hbox to 80pt{\em Free edges\hfill}]those that contain vertices
  that do not belong to the $n$-gon.
We call that latter vertices {\em free vertices\/}.
\item[\hbox to 80pt{\em Span edges\hfill}]all others.
\end{itemize}

To better discern the vertices in an MMP hypergraph with over 24
vertices and over 40 edges, we usually represent them by two
figures---one showing the $n$-gon with free edges and the other showing
the $n$-gon with span edges.

{\noindent {\em 42-24} \ (13) \  {\1
    3124,4VIU,UX97,7586,6WOd,dHBT,TRSM,MKJL,LcCb,bPGf,fgAe,eYQa,aZE3,
   9ABC,DEF8,GHIJ,NOPQ,WXYF,ecVD,gUSO,ZTN7,bWR2,fK63,eJ72.}}
Shown in Figs.~\ref{fig:even24-30a}..

{\noindent {\em 50-30}  \  (15) \ {\1
   3124,4DEF,Fm6i,ihbP,POQJ,JHIG,GoCj,jkKS,SRTU,UeLd,dl7W,WVNX,Xg8f,fnAZ,
ZYa3,5678,9ABC,KLMN,bcaM,TQFC,ecEB,lkPA,mdZO,mgRI,iYKH,njcW,jhg4,VHA4,oaR7,
oife.}} \quad Shown in Fig.~\ref{fig:even24-30b}.

{\noindent {\em 54-34} \ (16) \ {\1
    1234,4567,789A,ABCD,DEFG,GHIJ,JKLM,MNOP,PQRS,STUV,VWXY,YZab,bcde,
    efgh,hijk,klm1,3EQb,5DWl,7JYk,8Ubl,CHXe,DKTa,EMfm,OTck,2nLZ,3To9,6FLq,7Nen,Bonm,\break
    IQWn,Ipdi,LRlp,Srsj,Ugrq.}}  \quad Shown in Fig.~\ref{fig:54-34a}.

{\noindent {\em 60-40} \ (18) \ {\1
     3124,4576,6yau,uvVt,trqs,soTP,Pxh9,98AB,BpWM,MJLK,KicU,UwOl,lmnk,kjSH,
     HGIF,FCED,DYRf,feg3,,NOPM,QRSB,TUVW,XYZW,abcd,hiI2,odZS,pjc7,qLA6,wtbH,xvpg,yxnR,
rhSJ,vOEA,mYP6,ieCB,oneG,ncXA,ulhf,reaO,wpoD,slB4}}
\quad Shown in Fig.~\ref{fig:54-34b}.

To obtain these hypergraphs, we used a procedure that strips one edge at a
time.  For $n$ input hypergraphs each with $b$ edges, $n\cdot b$ output
hypergraphs, each with $n-1$ edges, were generated.  After
passing these output hypergraphs through several filters to eliminate
unconnected hypergraphs,
duplicates, non-KS (colorable) sets, and isomorphic sets, a smaller
number of hypergraphs usually resulted.  In order to keep the run time
feasible, we took a semi-random\footnote{We say ``semi-random'' because
while we chose random edges from each input hypergraph, we did not
choose random input hypergraphs.  This may have a small effect on
some of the inferred statistics; see the first paragraph of
\ref{sec:appstat}.} sample of the generated hypergraphs so that in
the end we would have approximately the same number $n$ of hypergraphs to
send to the next edge-stripping step.

The programs we used and their algorithms are described in Sec.~\ref{sec:alg}.
We used the program {\tt mmpstrip} to strip edges from
starting hypergraphs.  We adjusted the increment parameter of {\tt
mmpstrip} so that, after each edge removal and post-processing step, we
ended up with a sample of a desired size, say 50,000 hypergraphs.  We
ended up with 64 sample sets of up to 50,000 hypergraph each, with one
sample set for each hypergraph size from 12 through 75 edges.  For
12 edges and less, no KS sets have ever remained, probably because
they don't exist.

The complete processing of the samples sets of this size, including
finding all of the critical KS hypergraphs among them, took about 4 days
on a single CPU.  We ran 200 such jobs on a cluster, then combined the
results.  The random selection and the large sample space ensured that
we would have, with high probability, completely different samples on
successive program runs.  Except near the extreme edge sizes of 75 and
12 where the sample space is essentially exhausted, we never found a
duplicated hypergraph in our spot checking.

The overall iterative procedure we used is as follows.  We started with
the MMP hypergraph for the 60-75 KS set.
\begin{enumerate}
\item
We started with the 60-75 KS set and with  {\tt mmpstrip} obtained
75 new 60-74 sets---each with one less edge
than the original 60-75. Then we repeated the procedure
to obtain 2775 new 60-73 sets, etc.\ but not more than 50,000 hypergraphs
(or a million for some studies) to keep the run time reasonable.
We used {\tt mmpstrip}'s increment parameter $i$, which selects every $i$th
edge on average, to achieve this limit.
The increment parameter can be a non-integer
to better control the output size.  When it is greater than 1,
the edges can be selected either with uniform spacing or randomly;
the latter option was usually chosen.
\item A hypergraph is {\em connected} if there is a chain of zero or
more edges connecting every pair of edges.  Unconnected hypergraphs
were removed with {\tt mmpstrip}.
\item  Duplicate hypergraphs result when one edge is removed
at a time (rather than multiple edges combinatorially).
These duplicates were removed.
\item
  Isomorphic hypergraphs were removed with {\tt shortd}.
\item
  Colorable hypergraphs were filtered out with {\tt states01},
  leaving only non-colorable ones (i.e.\ KS sets).
\end{enumerate}
Typically, we first ran these steps on  a small
sample of the hypergraphs (a hundred or so) so that the increment
parameter for each {\tt mmpstrip} call could be estimated,
in order to end up with the same number of output hypergraphs as
input hypergraphs after each edge removal step.

We examined the final set of MMP hypergraphs obtained after each iteration
of the above process in order to determine which hypergraphs were critical,
using an option of the {\tt states01} program.  Any critical sets found
were collected for analysis.

The advantage of stripping one edge at a time then filtering at each
stage is that many fewer MMP hypergraphs had to be examined, because at
each step we consider only the non-isomorphic KS sets from the previous
step.  For example, from Fig.~\ref{fig:stat} (below), there are
$3.1\cdot 10^{20}$ hypergraphs with 28 edges, of which only $1.6\cdot
10^{13}$ are KS sets.  The 23 critical sets were were found by examining
a sample of only $6\cdot 10^7$ KS sets rather than $1.2\cdot 10^{15}$
starting MMP hypergraphs that would have been required, representing a
speedup factor of about 20 million.

\section{Algorithms}\label{sec:alg}

For the purpose of the KS theorem, the vertices of an MMP hypergraph are
interpreted as rays, i.e.\ 1-dim subspaces of a Hilbert space, each
specified by a representative (non-zero) vector in the subspace.  The
vertices on a single edge are assumed to be mutually orthogonal rays or
vectors.  In order for an MMP hypergraph to correspond to a KS set,
first there must exist an assignment of vectors to the vertices such
that the orthogonality conditions specified by the edges are satisfied.
Second, there must not exist an assignment (sometimes called a
``coloring'') of 0/1 (non-dispersive or classical) probability states to
the vertices such that each edge has exactly one vertex assigned to 1
and others assigned to 0.

For a given MMP hypergraph, we use two programs to confirm these two
conditions.  The first one, {\tt vectorfind}, attempts to find an
assignment of vectors to the vertices that meets the above requirement.
This program is described in Ref.~\cite{pmm-2-09}.  The second program,
{\tt states01}, determines whether or not a 0/1 coloring is possible
that meets the above requirement.  The algorithm used by {\tt states01}
is described in Ref.~\cite{pmmm03a}.  An additional option was added to
{\tt states01} to determine if a hypergraph is critical i.e.\ whether the
hypergraph is non-colorable but becomes colorable if any single edge is removed.

The 60-vertex, 75-edge MMP hypergraph based on the 600-cell described
above (which we refer to as 60-75) has been shown to be a KS set.\
\cite{aravind-600} However, it has redundancies (is not a critical set)
because we can remove edges from it and it will continue to be a KS
set.  The purpose of this study was to try to find subsets of the 60-75
hypergraph that are critical i.e.\ that are minimal in the sense that if
any one edge is removed, the subset is no longer a KS set.

While the program {\tt vectorfind} independently confirmed that 60-75
admits the necessary vector assignment, such an assignment remains valid
when a edge is removed.  Thus it is not necessary to run {\tt
vectorfind} on subsets of 60-75.  However, a non-colorable (KS) set
will eventually admit a coloring when enough edges are removed, and the
program {\tt states01} is used to test for this condition.

A basic method in our study was to start with the 60-75 hypergraph and
generate successive subsets, each with one or more edges stripped off
of the previous subset, then keep the ones that continued to admit no
coloring and discard the rest.  Of these, ones isomorphic to others were
also discarded.

The program {\tt mmpstrip} was used to generate subsets with edges
stripped off.  The user provides the number of edges $k$ to strip from
an input MMP hypergraph with $n$ edges, and the program will produce
all $\binom{n}{k}$ subsets with a simple combinatorial algorithm that
generates a sequence of subsets known as the ``banker's sequence.''
\cite{loughry-subset}
Partial output sets can be generated with start and end parameters.
By default, {\tt mmpstrip} will scan linearly through the edges to pick
every $i$th one when the increment parameter is $i$.  The program will
optionally randomise this edge selection, so that while a fraction
$1/i$ of edges is picked from each input hypergraph, which edges are
picked are random. The goal of this feature is to lessen the chance of
a biased selection due to a pattern that is repeated for every
hypergraph (such as removing only the first edge from each hypergraph).
It is hoped that the samples would thus provide a more
uniform representation of the search space.

Optionally, {\tt mmpstrip} can take truly random samples with
replacement (for a given number of edges) from the starting 60-75 MMP
hypergraph (in contrast to the semi-random method of the previous
paragraph).  This mode was used to verify or improve some of the
statistical estimates in Fig.~\ref{fig:stat}.  A cryptographic hash of
the time of day, process ID, and CPU time is used as the seed for the
pseudo-random number generator.  The seed may also be provided by the
user in order to repeat a result.

The {\tt mmpstrip} program will optionally suppress MMP hypergraphs
that are not connected, such as those with isolated edges or two
unconnected sections, since these are of no interest.  The
output lines are by default renormalized (assigned a canonical vertex
naming), so that there are no gaps in the vertex naming as is required by
some other MMP processing programs.

In order to detect isomorphic hypergraphs, one of two programs was used.
For testing small sets of hypergraphs, we used the program {\tt
subgraph} described in Ref.~\cite{pmm-2-09}, which has the advantage of
displaying the isomorphism mapping for manual verification.  For a large
number of hypergraphs, we used Brendan McKay's program {\tt shortd},
which has a much faster run time.


Program {\tt longest} singles out longest loops from the list of all
possible loops (which is the output of the program {\tt loopbig}).
Programs {\tt parse} or \verb|parse_all| then ``write'' a
program or programs in the vector graphics language {\tt Asymptote}
for drawing a chosen hypergraph or all hypergraphs, respectively.

The longest loop of each hypergraph is drawn as $n$--sided regular
(equilateral and equiangular) polygon, where $n$ is the number of
edges in the loop.
By default, free vertices, i.\,e.\ vertices that are not on the
loop, are placed inside the polygon, off--centre, on vertical
lines, with not more than 4 vertices on one line, but the user can change
options for their placement.
Edges contained in the longest loop are drawn as straight lines,
while other edges are drawn as B\'ezier curves (specifically,
{\tt Asymptote} is based on Donald Knuth's
{\footnotesize\sf METAFONT}).  The user can interactively
change the ``tension'' of
the curve and the amount of ``curl'' at its endpoints,
in order to interactively control and change the shape of each
edge line and unambiguously discern vertices that share an edge.


\section{Sample Space Statistics}\label{sec:stat}

There are $3.8\cdot 10^{22}$ possible subsets of the 60-75 set
(disregarding any symmetry) and, among them, approximately $7.5\cdot
10^{17}$ KS sets.  An exhaustive search for critical KS sets was
not feasible for the present survey, but it may become feasible in
the future, possibly requiring a year or more on a large computer
cluster.

For our survey, we searched a total of around $10^{10}$ KS sets,
randomly chosen for a given edge size, to find the critical KS sets
among them.  We then performed a statistical analysis to estimate the
total number of critical KS sets that would be found by an exhaustive
search.  The final result is that we can expect a total of $4.3\cdot
10^{12}$ non-isomorphic critical sets, with a 95\% confidence interval
between $4.0\cdot 10^{12}$ and $4.6\cdot 10^{12}$ based on the
statistical model we used.

If an exhaustive search is performed, it is possible to store the
complete set of non-isomorphic critical sets with current technology.
Without compression, each critical set (in MMP hypergraph notation)
requires an average of about 260 bytes, thus requiring $260\cdot 4.3\cdot
10^{12}=1.1\cdot 10^{15}$ bytes (1.1 petabytes) of storage.  This could
probably be reduced considerably with data compression techniques.

\begin{figure}[hbt]
\includegraphics[width=0.98\textwidth]{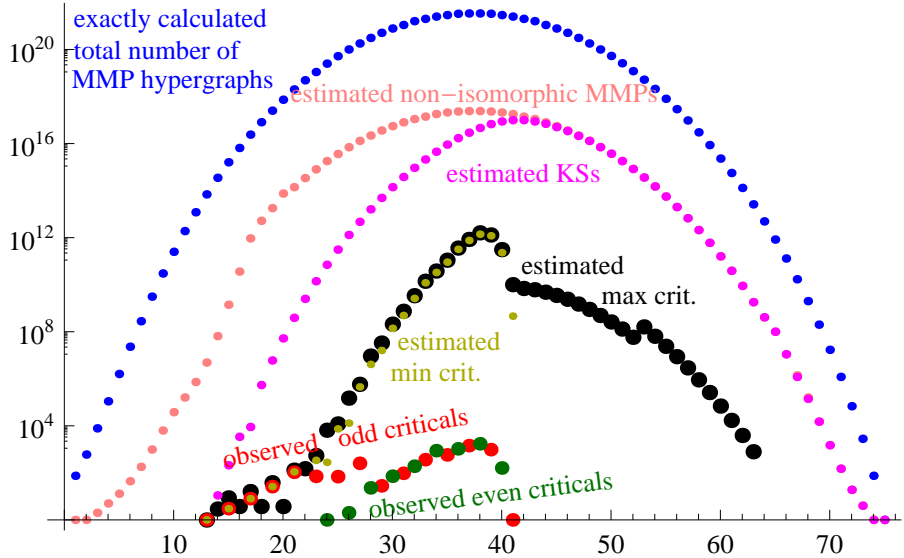}
\caption{\label{fig:stat}Overall statistics calculated for
subsets of 60-75 given on a logarithmic scale.
The samples (for 13-63 edges; 64-75 search was exhaustive)
contained on average $3\cdot10^8$
MMP hypergraphs. ``Observed odd (even) criticals'' refer to
odd (even) numbers of {\em critical\/} KS sets. The sudden jump
in the ``estimated max  crit.'' plot at 53 edges
is caused by a change in the sample size, as explained in
the text.  The ``estimated min crit.'' points are not plotted when
they are zero. The ``estimated max  crit.''\ for more than 63 edges
are not shown since the exhaustive search found no criticals there.}
\end{figure}

The plots of Fig.~\ref{fig:stat} provide an overview of the subsets of
60-75, broken down by the number of edges.  These plots are intended to
provide a guideline for estimating the work that would be required for an
exhaustive search for a particular number of edges or range of them.

The details of the statistical methods we used, as well as a more detailed
description of Fig.~\ref{fig:stat}, are given in \ref{sec:appstat}.

In our survey, no critical sets were observed in KS sets with 42 or more
edges.  In these cases, the ``estimated max crit.'' in
Fig.~\ref{fig:stat} has little to do with the number of critical sets (if
any) that actually exist in that edge range.  Instead, it could be
interpreted as ``zero with statistical noise'' and is primarily a
function of the number of samples we took and the search space size for
the particular edge size.  It simply indicates that, based on a
Bernoulli trial probability model, it is unlikely (with
95\% confidence) that there are {\em more} critical sets than ``estimated
max crit.''  The sudden jump between 52 and 53 edges is due to the fact
that we changed the number of samples from $5.3\cdot 10^7$ per edge size
to $8.6\cdot 10^6$.  If an exhaustive search shows that there are no
critical sets at all above 41 edges, that will be completely consistent
with the ``estimated max crit.'' bound.  In fact, we would conjecture
that the actual number of critical sets will be identically zero very
soon after 41 edges (see the last paragraph of \ref{sec:appstat}).

\section{Conclusions}
\label{sec:con}

Kochen-Specker (KS) sets and setups proposed, designed, and
experimentally carried out so far were either  3-, 4-, 8-, \dots
dimensional KS sets (Peres', Cabello's, etc.) or the Mermin set.
They aim at finding particular valuation of the KS observables
that prove the quantum contextuality and disprove any noncontextual
classical valuations of those observables. Our aim is to make KS
sets independent of a particular choice of either vectors or
observables so as to make them suitable for building quantum gates
within a would-be quantum circuit.

For this application, we should have a choice of gates of different
sizes, that is, consisting of sufficiently many vectors and sufficiently
many gates for a chosen number of vectors, and this is what we
achieved in the previous sections.  We generated a large number of
4-dim critical non-redundant non-isomorphic KS sets
with 26 to 60 vectors based on the 600-cell (the 4-dimensional analog
of the icosahedron).  ``Critical'' means that no orthogonal tetrads can
be removed without causing the KS contradiction to disappear.
In other words, they represent a KS setup that has no experimental
redundancy.

The generation was achieved by algorithms and computer programs
described in Sec.~\ref{sec:alg}, with which we found the critical sets
summarized in Fig.~\ref{fig:stat}. Previously, only two 30-15
critical KS sets were found by Waegell and Aravind \cite{aravind10}
as well as a new third class. In this work, we have extended our
previous study of up to 19 edges \cite{mp-nm-pka-mw-11} to sample
the entire range of all 60-75 subsets. The number of critical
sets with the lowest number of edges,
13 through 19, appears comparatively small, and it is feasible to
find all of them with an exhaustive search. Beyond that, the number
of non-isomorphic classes becomes very large and attempting their
generation now would take too many CPU months on the grid. In
Sec.~\ref{sec:stat} and \ref{sec:appstat},
we give the detailed statistical estimates of the
total critical sets that exist based on our samples.  
The statistical techniques we used are general-purpose 
and can be useful for any similar experiment in which 
an exhaustive enumeration of outcomes is not feasible.

We stress here that critical sets obtained by a 
future exhaustive generation might well be far less 
numerous then their statistical estimates given above. 
That would be yet another proof of how different   
quantum sets are from classical ones and how cautious we 
should be when interpreting classical statistical methods 
applied to quantum data. This is also the reason why we 
have not and could not have made any realistic predictions 
on numerosity and existence of critical sets that might 
be observed in the future but have not been observed so far.
All we can say right now is that the predicability of the total
number of KS sets is in good agreement with the data we 
so far obtained by exhaustive generation. For example, by 
exhaustive generation of KS sets with 63, 64, 65, and 66 
edges we obtain $1.8\cdot 10^9$, $4.1\cdot 10^8$, 
$1.0\cdot 10^8$, and $1.1\cdot 10^7$ sets versus 
estimated $1.8\cdot 10^9$, $3.4\cdot 10^8$, 
$5.7\cdot 10^7$, and $8.8\cdot 10^6$, respectively.

The main theoretical results of our generation are that
\begin{itemize}
\item the 24-24 and the 60-75 classes are disjoint (in the sense
that the biggest set of the 24-24 class is the single Peres' 24-24
set and the smallest set from the 60-75 class is the 26-13 one;
\item the maximal loop of all sets from the 24-24 class is always
a hexagon while the maximal loops of the sets from the 60-75
class grow (form at least an octagon) as the number of vectors and
edges increase (see Figs.~\ref{fig:30-15a}-\ref{fig:54-34b});
\item there is an unexpectedly large and rich universe with an
estimated \mbox{$4.3\cdot 10^{12}$} non-isomorphic critical sets
inside of the 60-75;
\item in \cite{bdm-ndm-mp-fresl-jmp-10} we found that one of the
known 3-dim KS sets passes a series of equations that hold in any
Hilbert space---the so-called orthoarguesian equations. We have
not found any such KS set in the 60-75 class so far. Both results
show that orthogonality of vectors does not suffice for a complete
Hilbert space description of KS sets---the relations between nonorthogonal
vectors play an essential in such a description. This will prove
essential for a proper description of quantum gates using KS sets;
\item there is only one KS set with 24 vectors (vertices) and
24 tetrads (edges), and it contains all KS sets from the 24-24
class with the chosen values of vector components.~\cite{pmm-2-09}
In contrast to this, there are many non-isomorphic KS sets
with 60 vectors and 60 tetrads which contain many non-isomorphic
KS subsets each.
\end{itemize}

Another open question is to find physical and geometrical reasons for
having only hexagon maximal loops in the 24-24 class and for
having particular octagons, nonagons, decagons, etc., in the
60-75 class.

\appendix
\section{Samples of KS hypergraphs with even number of edges}
\label{sec:app}

Here we give samples of KS hypergraphs of each kind that we listed
in Table  \ref{table:even} and did not give in
Figs.~\ref{fig:30-15a}--\ref{fig:54-34b} and in
Sec.~\ref{sec:res}.

Using our programs  {\tt longest} and  {\tt loopbig}, we can instantly
determine the following structural features. Let us take
the  {\em 45-26} hypergraph below. The program {\tt longest} shows that its
biggest loop is a 12-gon. The program {\tt loopbig} gives 26 instances of
its 12-gon representation, the first one of which is
{\12134,4YZE,\break EFGD,DfKN,NPQO,OeUd,dacb,bWL7,7586,6jMT,ThgV,ViR2.
 9.A.B.C. H.I.G*8* J.K*L*M*\break R*S.Q*C. T*U*S.F* V*W*X.P* c*X.M*3* f*Z*R*7* g*e*I.3* i*j*Y*N* h*a*Y*H. a*B.6*2* e*Y*L*A. d*V*K*9.}
(The edges are the same as in {\em 45-26} below, only in a 
different order. Also the vertices within an edge are mostly 
in a different order. Actually, all 26 instances are just 
different 12-gon arrangements of {\em 45-26} below.)  
The edges {\11234--ViR2} are {\em polygon edges} (see
Sec.~\ref{sec:res}); vertices followed by ``.''  are
{\em free vertices}; edges containing {\em free vertices}
are {\em free edges}; vertices followed by ``{\1*}''  are {\em polygon  vertices};
edges containing only  {\em polygon  vertices} followed by ``{\1*}''s
are {\em span edges}; (in other words {\em span edges} are edges which
are not {\em polygon edges} and which do not contain {\em free vertices}).
Our script based on {\tt Asymptote} draws 26 figures of {\em 45-26}
with 12-gons. Once the figures are drawn, the user can
assign any ASCII symbol desired to any vertex. Also, by utilizing our
program {\tt vectorfind} she/he can ascribe vectors to vertices.
{\11,2,\dots,i,j} $\to$ \{$\tau$,0,$\overline{1}$,$\overline{\kappa}$\},$\>$\{0,1,0,0\},
\{$\kappa$,0,$\tau$,$\overline{1}$\},$\>$\{1,0,$\kappa$,$\tau$\},$\>$\{$\kappa$,$\tau$,1,0\},$\>$
\{$\overline{1}$,0,$\kappa$,$\tau$\},$\>$\{$\overline{\tau}$,$\kappa$,0,$\overline{1}$\},$\>$
\{0,$\overline{1}$,$\tau$,$\overline{\kappa}$\},$\>$\{0,$\tau$,$\kappa$,1\},$\>$
\{1,$\overline{1}$,1,1\},$\>$\{$\overline{\kappa}$,0,$\tau$,$\overline{1}$\},$\>$
\{$\tau$,$\kappa$,0,$\overline{1}$\},$\>$\{$\overline{1}$,1,1,$\overline{1}$\},\break 
\{$\overline{1}$,1,$\overline{1}$,1\},$\>$\{$\overline{1}$,$\overline{1}$,1,1\},$\>$
\{1,1,1,1\},$\>$\{$\overline{\kappa}$,$\overline{1}$,0,$\tau$\},$\>$
\{$\tau$,$\overline{1}$,$\overline{\kappa}$,0\},$\>$
\{$\kappa$,$\overline{\tau}$,$\overline{1}$,0\},$\>$
\{0,$\overline{1}$,$\tau$,$\kappa$\},$\>$
\{$\overline{1}$,0,$\overline{\kappa}$,$\tau$\},\break \{$\tau$,$\kappa$,0,1\},$\>$
\{$\tau$,1,$\kappa$,0\},$\>$\{$\kappa$,0,$\overline{\tau}$,$\overline{1}$\},$\>$
\{0,$\kappa$,$\overline{1}$,$\tau$\},$\>$
\{$\overline{1}$,$\tau$,0,$\overline{\kappa}$\},$\>$
\{0,0,1,0\},$\>$\{$\kappa$,1,0,$\tau$\},$\>$\{0,$\overline{1}$,$\overline{\tau}$,$\kappa$\},$\>$
\{$\tau$,$\overline{1}$,$\kappa$,0\},$\>$\{1,0,0,0\},$\>$\{0,1,$\tau$,$\kappa$\},$\>$
\{0,$\tau$,$\overline{\kappa}$,$\overline{1}$\},$\>$\{$\overline{1}$,$\kappa$,$\tau$,0\},$\>$
\{$\overline{1}$,$\overline{\tau}$,0,$\kappa$\},$\>$\{$\tau$,0,1,$\kappa$\},$\>$
\{$\kappa$,$\tau$,$\overline{1}$,0\}, \{$\overline{1}$,1,1,1\},$\>$
\{0,$\overline{\kappa}$,$\overline{1}$,$\tau$\},$\>$\{1,$\tau$,0,$\kappa$\},$\>$
\{$\kappa$,$\overline{1}$,0,$\overline{\tau}$\},$\>$
\{0,$\kappa$,$\overline{1}$,$\overline{\tau}$\},$\>$\{0,$\overline{\tau}$,$\kappa$,$\overline{1}$\},
\{0,0,0,1\},$\>$\{$\overline{\kappa}$,$\tau$,$\overline{1}$,0\}, where
$\tau=(\sqrt{5}+1)/2$ and $\kappa=1/\tau$; a bar over a
number indicates its negative.

\smallskip
{\baselineskip=10pt
{\noindent {\em 45-26} \   {\1
1234,5678,9ABC,DEFG,HIG8,JKLM,NOPQ,RSQC,TUSF,VWXP,YZE4,abcd,edUO,\break
cXM3,fZR7,bWL7,geI3,fNKD,hgVT,ijYN,haYH,jTM6,aB62,iVR2,eYLA,dVK9.}}

{\noindent {\em 46-28} \  {\1
1234,5678,9AB8,CDEF,GHIJ,KLMN,OPQR,STR4,UVNF,WXYZ,abZT,YQME,cdYB,\break
efb7,gVSJ,hiPA,jfOI,idHD,aLIC,ieXN,jiS6,kgM5,khcG,kbU9,hL73,cON2,WSL9,gZOD.}}

{\noindent {\em 47-28} \  {\1
1234,5674,89AB,CDEF,GHIJ,KLJ7,MNOB,PQR3,STUO,VWU6,XYI5,ZabL,cdYL,efbF,
gaHA,hWG9,iTGE,jkZ2,lkNE,kdUR,lfXK,hXD1,XVQ8,jfPA,jcMC,ecSQ,khge,iaXM.}}

{\noindent {\em 48-28} \  {\1
1234,5678,9ABC,DEC8,FGHI,JKLM,NOME,PQRB,STIA,UVWX,YXRL,Zab4,cdW3,efgT,
hdbK,ijcJ,gVQH,kfJG,lhF7,jeb6,iUPD,faU9,YS72,mlZT,lQO3,kdYN,mUN6,iZYH.}}

{\noindent {\em 49-28} \  {\1
1234,5678,9ABC,DEFG,HIJ8,KLG7,MNOP,QRSP,TUVJ,WLC4,XYZV,abS7,cdeb,fgUO,
heZN,ijdI,kgaF,lcYM,jYRE,iXQG,fHGB,mfdW,mkTN,nhR3,nigA,lhFC,cTG3,mYA7.}}

{\noindent {\em 49-30} \  {\1
1234,5678,9ABC,DEFG,HIJK,LMN8,OPQR,STUK,VWRC,XYZW,abcQ,deJ4,fghe,ijU3,
kjQG,ZTNB,khA7,gYS8,lkV2,mcMF,mif6,lidE,nfXD,mPJB,gOI3,bX73,ndcC,lgaB,nkNI,lXMK.}}

{\noindent {\em 51-30} \  {\1
1234,5678,9AB8,CDEF,GHIJ,KLMN,ONJ4,PQRS,TUB3,VWIA,XSOF,YZE7,aHD6,bcZX,
defW,ghiM,cWR2,jkL9,liR5,mbUC,nkfb,onha,pgeX,pojG,mhdQ,oYP3,pmVK,nlTK,kgYH,ljd4.}}

{\noindent {\em 52-30}  {\1
1234,5678,9ABC,DEF4,GHIJ,KLMN,OPQ8,RSQJ,TUVI,WXYZ,abcJ,NHC3,defg,hijc,\break
kgbB,lkZ7,mYMJ,njX4,fSNF,oWVA,piPA,qeVE,onml,qpna,ohdQ,pfYU,liLE,UQLB,qhN7,ngI8.
}}

{\noindent {\em 53-30} \  {\1
1234,5678,9ABC,DEFG,HIJK,LMNO,PQRS,TUVW,XYZW,abZS,cdVG,eRC4,fghY,ihK3,
gdbJ,jkeI,lQOF,mki8,nopJ,qpjE,qhP7,roX7,nmUQ,rjcN,rfUB,pliT,paMB,mYNC,MIG7,lbC7.}}

{\noindent {\em 51-32}   {\1
1234,5678,9AB4,CDEF,GHFB,IJKL,MNOP,QRPL,STUH,VWXR,YZUQ,abcO,dZXE,\break 
eTK8,fgJA,hig3,jiN7,iSD9,khcY,lXMK,mnWA,opYI,ndc6,ljfC,mjZ2,lbS5,pV62,piaG,ogeb,keWG,\break 
onPC,kSP2.}}

{\noindent {\em 52-32}  {\1
1234,5674,89AB,CDE7,FGHB,IJKL,MLA3,NOP2,QRST,UVWX,YZaX,bcWT,dePK,fgSJ,\break 
hiaH,jkOG,lmW9,nmgN,oeVR,kife,pkZQ,pcE3,onHE,qjRM,qfbY,hbND,qp96,laR4,UJGD,qndU,\break 
dZB7,RNIB.}}

{\noindent {\em 53-32}  {\1
1234,5678,9A84,BCDE,FGHI,JKLM,NOPQ,RST3,UVW7,XYIA,Zabc,defM,ghYL,ijcT,\break 
kWQ2,ljfH,mhbV,nmiP,eSGE,oaUK,pgZR,qohO,plNK,ondI,ZXE2,rfUD,gdC6,qiD9,rpmk,qkJG,\break 
cNG5,mHC3.}}

{\noindent {\em 54-32}  {\1
1234,5678,9ABC,DEFG,HIJC,KLMN,OPQR,STUC,VWXY,URN4,Zabc,def3,ghcQ,ihMJ,\break 
jkB8,lifY,mkgT,nopA,qrlT,rpjP,ebX8,rhdG,qoeJ,maPF,naNI,sbLE,som7,oZYG,nWTE,rVN7,\break 
fQEB,YPLC.}}

{\noindent {\em 55-32} {\1
1234,5678,9ABC,DEFC,GHIJ,KLMF,NOPQ,RSTU,VWXY,ZabE,cde8,fghe,ijeQ,klmU,\break 
mhbJ,nopB,qrpT,sjT4,todY,naMI,SHA3,tlP7,ngR7,rGF7,qZXO,tsZL,eWM3,fVLA,nmiV,dNF4,\break 
qkdA,mOC8.}}

{\noindent {\em 53-34} \ \ \  {\1
1234,5674,89AB,CDEF,GHIJ,KLMN,OPQR,STUV,WVRB,XYUF,Zabc,decN,fgYJ,hijA,
kjbE,lmQF,ePI3,aWMI,niXL,oZRH,phcT,pmH7,qkLB,pgKE,qhI6,rnmf,ncO2,mdSA,OJDA,roke,\break
ogS6,lLJ4,raU4,nIE9.}}

{\noindent {\em 55-34} \  \  \  {\1
1234,5674,89AB,CDEF,GHIJ,KLJ7,MNOP,QRST,UVWT,XYZa,bcaW,dec6,fghI,ihPB,
jkgS,kZVL,lmRO,noi3,pqYI,rNF3,qoeE,mbDA,sljc,rplU,sfXQ,kbH2,tsqN,ndYO,tU97,nfLA,\break
reSJ,XUPH,siJD,kOE9.}}

{\noindent {\em 56-34} \  {\1
1234,5674,89AB,CDEF,GHIJ,KLMN,OPQR,STUJ,VWXB,YURN,Zabc,defT,ghMF,ijQE,
klmA,nLI9,opnc,pmhY,qjbH,oKD8,olX7,rsfa,iWS3,tsWM,utZP,urmJ,neO6,uqog,dHF3,skH6,\break
qYV4,laIE,ieZY,fPD4.}}

{\noindent {\em 57-34} \  \ {\1
1234,5674,89AB,CDEF,GHIJ,KLMN,OPQB,RSTU,VWXY,Zabc,defc,ghb7,ijkY,lmnU,
ofT3,pqSA,rstR,qnXQ,uvmN,pkhN,ljgP,ieMA,tolJ,vdWI,aQJF,ueVP,ZPLE,sPI4,RNHD,voiZ,\break
rnL3,bYDB,rpVF,dRQ7.}}

{\noindent {\em 58-34} \  {\1
1234,5678,9ABC,DEF8,GHIJ,KLMN,OPQN,RSTU,VWXY,ZaMJ,bYU4,cdef,ghia,jkiQ,
lmnT,onP8,phfL,qoeS,rROI,hbHC,srp3,qpmX,tuvo,wvdJ,wqb7,rkdK,jcWB,uscb,trlB,ukVT,\break
lbZF,gVIF,vpjF,qgNB.}}

{\noindent {\em 59-34} \  \ {\1
1234,5678,9ABC,DEFG,HIJK,LMNO,PQR8,STUV,WXYZ,abcZ,defg,higO,jklm,nopm,
qric,slbR,tukQ,uV74,vuYN,srf3,paN3,wthC,jeKB,xwvj,qodX,xqUQ,jVMG,UJF3,nIC8,wWTH,\break
cTE8,snWM,wdRF,oTOB.}}

{\noindent {\em 55-36} \  {\1
1234,5678,9AB4,CDEF,GHIJ,KLMF,NOP3,QRPM,STUJ,VWOB,XYZa,bcdA,efgh,ijkd,
lmkU,haNJ,nocW,pREB,qrpj,mbZ2,ojJD,srM8,qgb7,tlaW,WTF7,qnQI,kYHE,sZIB,fXOD,iL62,\break
tfI6,qSOL,leLA,neE8,rWH2,kfM4.}}

{\noindent {\em 56-36} \  {\1
1234,5674,89AB,CDEF,GHIJ,KLMF,NOPQ,RST3,UVWX,YZaX,bcdW,efgB,hVE7,ijgQ,
kTPM,lmnd,oQJD,pqrs,faC2,tsnI,kjcA,tZOA,uib6,mYS9,uonf,rliR,sUQL,qXPH,poZ4,rG97,\break
mkeU,hRIB,qmh2,mOK6,ocRK,bPC9.}}

{\noindent {\em 57-36} \  \ {\1
1234,5674,89A3,BCDE,FGHI,JKLM,NOPQ,RSTU,VWXY,ZaYQ,bcdI,efE2,ghM3,ijkf,
lhda,mecP,nOL7,oXU6,pqbA,kNHD,rsqZ,pnfT,cWSK,ZJGE,tunl,vsVO,ljUG,vgTD,umiR,oiOI,\break
rmh6,iWCA,vtA6,usH4,mYD9,qlKD.}}

{\noindent {\em 58-36} \  {\1
1234,5674,89AB,CDEF,GHI3,JKLM,NOPQ,RSTM,UVB7,WXYI,ZaYT,bcaQ,dcSF,efgh,
ijkR,lmhP,ngXL,nmZH,opWV,qplK,rpaA,sJE4,trkG,tnD9,usoZ,vfbI,qjf6,wXO6,wtse,vtld,\break
kbVC,uhUM,wpiF,ujdN,kh84,pnN4.}}

{\noindent {\em 59-36} \  \ {\1
1234,5674,89AB,CDEF,GHB7,IJKL,MNOP,QRPF,STUV,WXYA,ZabL,cdef,ghij,kfbY,
kVR6,lKH3,mjO9,nopX,qrlU,siXT,tule,vwpS,odRK,xusJ,xwc6,thYG,daOE,wrm2,iND4,uZSA,\break
xqF9,rnfN,vtNL,rgEA,phPJ,XLF2.}}

{\noindent {\em 60-36} \  {\1
1234,5678,9AB8,CDEF,GHIJ,KLMN,OPQJ,RSTU,VWXN,YZab,cdeB,fgX7,hgeM,ijQF,
kljd,mnoN,phbI,qrif,saEB,tolH,uvn4,srUL,wMD3,xmkf,wroZ,vcbT,ywjS,hWSP,ukYL,kVTJ,\break
yvHE,uqP8,xUH8,tfbD,poQB,fSB4.}}

{\noindent {\em 57-38} \  {\1
1234,5674,89AB,CDE7,FGHI,JKLM,NOPB,QRS3,TUVW,XYZa,bcde,fgPM,heWE,ijVS,
klRI,mnol,podH,qrU6,sraP,toOL,ukaK,tgZG,cKD3,vspQ,vueO,vTJ7,vqgA,pkjh,nibP,jfcY,\break
rmjG,nY94,viXI,tiE2,tkT9,qYHE,mQEB,cUIB.}}

{\noindent {\em 58-38} \  \ {\1
1234,5674,89AB,CDEB,FGHI,JKL7,MNOP,QRST,UVWX,YZaA,bcLE,defg,hia3,jkXP,
lmZK,ngWT,opnI,qrkH,rpSB,sfRO,tjRK,sYVL,umUG,vcPF,wm96,qeUD,okid,udbQ,wdYN,vpf2,\break
wnhc,qnMK,utra,viW9,sliD,wtD2,QPD7,usn4.}}

{\noindent {\em 59-38} \  {\1
1234,5674,89A3,BCD7,EFGH,IJKL,MNOL,PQRS,TUVW,XYZW,abZ2,cdeb,fgK6,hijA,
aVSH,klJ9,mnoG,pID8,qrYC,stO1,oeRO,ujgF,vuol,wutU,xkid,wnZJ,mjcX,rfdQ,rhaN,xwPC,\break
vcNE,qpkR,wpfM,vpV5,rmT8,tmkB,sWPF,vsiK.}}

{\noindent {\em 60-38} \  \ {\1
1234,5674,89A7,BCDE,FGHI,JKLM,NOPE,QRST,UVWT,XYZa,bcde,fghA,eWMI,ijkl,
mnlV,opha,qrgU,spnD,tdL3,trSP,cZRK,uveO,qkYI,jRHD,wrZ9,xiQ4,vtXV,mJC7,ywN6,yxtf,\break
pibN,ysKG,wjhd,romG,upT2,xumY,yqC2,viGA.}}

{\noindent {\em 59-40} \  {\1
1234,5678,9ABC,DEF8,GHI4,JKLM,NOPI,QRSC,TUVW,XYZH,abMB,cWPC,debZ,fgh7,
ijkF,lmnk,opje,qnaV,rsmU,tjU3,lhYS,uSE2,sqdA,kgXR,upfO,vTQN,qfL3,wnNK,vuJG,wfbF,\break
xicK,tgcE,vqoX,xura,xlTA,vcb6,spcY,rgeN,leG8,XUK8.}}

}

\section{Details for Sample Space Statistics}\label{sec:appstat}

The plots of Fig.~\ref{fig:stat} provide an overview of the subsets of 60-75.
Because they were determined by statistical inference from small samples
of this space, most of the numbers are approximate.  As a practical
matter, some of the sample sets, or portions of them, were obtained with
the more efficient semi-random method mentioned in the first footnote in
Section~\ref{sec:res}, which has an effect.\footnote{To test this
effect, we used non-isomorphic MMPs with 67 edges, where the actual
count is known.  Using semi-random sampling, a value of $1.4\cdot 10^6$
was estimated, compared to the actual count of $1.2\cdot 10^6$.  This is
apparently due to the more uniform sample provided by the semi-random
method, leading to the overcount.  Using true random sampling,
the estimate was very close to the actual $1.2\cdot 10^6$, as we
describe below.} Overall, the numbers should be trusted only to within
an order of magnitude or so.  The plots are intended to provide a rough
guideline for planning future work, such as an exhaustive search of
certain ranges, and for that purpose it should be adequate.

Several techniques, which we describe below, were used to obtain the
values for the plots. The total number of MMP hypergraphs is simply
$\binom{75}{b}=\frac{75!}{b!(75-b)!}$,
where $b$ is the number of edges given at the abscissa.

The {\tt mmpstrip} program was used to identify and remove unconnected
hypergraphs.  We do not include the resulting numbers of MMP hypergraphs
in Fig.~{fig:stat} but briefly describe them as follows.
For 1 through 4 edges, the number of unconnected MMP
hypergraphs are exactly 0, 2175, 59725, and 1101450.
For 67--75 edges there are exactly 0.  For the rest, we used
samples of $10^6$ MMP hypergraphs for each number of edges.
For 47--66 edges, no unconnected hypergraphs were observed.
For 5--46 edges, the number of unconnected hypergraphs (estimated
from the ones observed in the sample)
decreases to zero as a percentage the total number of MMP hypergraphs,
from  $1.56\cdot 10^{7}$ (out of $1.73\cdot 10^{7}$ total) for
5 edges to $1\cdot 10^{15}$ (out of $5.1\cdot 10^{20}$ total)
for 46 edges.

To calculate the the number of non-isomorphic MMP hypergraphs,
unconnected hypergraphs were
discarded and the rest passed through the {\tt shortd} program, which
filters isomorphic hypergraphs, keeping only one canonical representative
from each isomorphism class.  For small and large numbers of edges,
exhaustive generation of all MMPs yielded exact values.  For 1--4 edges
there are 1, 1, 2, and 5 (connected) isomorphism classes; for 67--75
edges, there are 1183189, 141314, 15014, 1463, 154, 19, 4, 1, and 1.
For the other edge sizes, the number of isomorphism classes was estimated
from a sample.  Finding this estimate is called the ``coupon collector's
problem,'' \cite{finkelstein} and the maximum likelihood estimator is
the smallest integer $j\ge c$ such that
\begin{align}
  \frac{j+1}{j+1-c}\left(\frac{j}{j+1}\right)^n<1,   \label{eq:coupon}
\end{align}
where $n$ is the number of samples (with replacement) and $c$ is the
observed number of isomorphism classes in the sample.  For example, we
observed $c=516604$ isomorphism classes in a random sample of $n=545961$
13-edge hypergraphs.  The criteria of Eq.~(\ref{eq:coupon}) yields
$j=4893025 \approx 4.9\cdot 10^{6}$, which is the point shown
for 13 edges in the non-isomorphic MMP hypergraphs plot of Fig.~\ref{fig:stat}.
We mention that in our
implementation, we expressed Eq.~(\ref{eq:coupon}) as
$\log(j+1)-\log(j+1-c)+n (\log j - \log(j+1))<0$ and determined $j$ with
a binary search method.  Because the computation involves the
subtraction of almost-equal terms, high-precision floating-point
operations are necessary.  For the calculations of Fig.~\ref{fig:stat},
Eq.~(\ref{eq:coupon}) gave incorrect answers with less than 35
significant digits, and we used 100 significant digits for robustness.

As a rough check of the statistical model used by the coupon
collector's problem, 10 random samples of 50000 67-edge MMP hypergraphs
yielded from 48900 to 48975 isomorphism classes, corresponding to
predictions of 1119613 to 1202764 total classes by
Eq.~(\ref{eq:coupon}).  This compares to the actual number of 1183189
classes obtained by exhaustive generation of MMP hypergraphs.

To estimate the KSs in Fig.~\ref{fig:stat}, KS sets were
identified using the {\tt states01} program.  For small numbers
of edges ($\le 12$), we never observed a KS set.  For large numbers
of edges ($\ge 63$), we never observed a non-KS set, so for them
the two plots coincide.  For those in between, we took a random sample of
non-isomorphic hypergraphs for each edge size and plotted
 the fraction of observed KS sets times the
estimated non-isomorphic MMP hypergraphs.

We show the number of isomorphically unique critical
hypergraphs we observed, as identified by the {\tt -c} (``critical'')
option of the {\tt states01} program,
in the ``observed odd criticals'' and ``observed even critical'' plots
of Fig.~\ref{fig:stat}.  We include these to show
the actual currently known (not estimated) number of critical sets.  It is
not, however, intended to convey the distribution of critical
hypergraphs vs.\ edge size; for that purpose, the estimated
maximum number of critical sets in Fig.~\ref{fig:stat} should be
used.\footnote{The ``observed odd [even] criticals'' in Fig.~\ref{fig:stat}
are not directly related to the distribution of critical sets
vs.\ edge size because we used varying sample sizes.
For example, the 879 critical sets with 34 edges were observed in
$1.1\cdot 10^8$ KS samples whereas the 580 critical sets with 35
edges were observed in only $5.28\cdot 10^7$ KS samples.  Since the
number of observed critical sets grows with the number of
samples, it is likely that the actual number of critical sets with 35
edges---that would be obtained with an exhaustive search---is larger,
not smaller, than the number with 34 edges.}

In the range of 12 through 62 edges, the ``estimated max crit.''
plot shows the upper 95\% confidence limit derived from Bernoulli
trial probabilities,
based on the model of sampling with replacement from a search space where the
{\em a priori} probability is unknown.
\cite{probability} If $K$ is the total number of KSs (from the
``estimated KSs'' plot), $n$ is
the sample size (with replacement) of random KS sets, and $m$ is the
observed number of critical sets, then the lower 95\% confidence level is
\cite[Eq.~(1)]{probability}
\begin{align}
    K \cdot I^{-1}_{\frac{1}{2}(1-0.95)}(m+1,n-m+1)  \label{eq:lowerconfidence}
\end{align}
and the upper 95\% confidence level
 is \cite[Eq.~(3)]{probability}
\begin{align}
    K \cdot I^{-1}_{\frac{1}{2}(1+0.95)}(m+1,n-m+1)
\end{align}
where $I^{-1}$ is the inverse regularized incomplete beta function.
For example, for the 35-edge case, $K=9.0\cdot 10^{15}$,
$n=52800000$, and $m=580$.  Thus for upper 95\% confidence level we have
$K \cdot I^{-1}_{\frac{1}{2}(1+0.95)}(m+1,n-m+1)
= K \cdot I^{-1}_{0.975}(581,52799421)\approx K \cdot 0.0000119163
\approx 1.1 \cdot 10^{11}$.  This is the value in the
``estimated max crit.'' plot for 35 edges.

The ``estimated min crit.'' plot shows either the the lower 95\%
confidence limit from Eq.~(\ref{eq:lowerconfidence}) or zero (in which
case we omit the ``estimated min crit.'' point from the plot since it is
outside the logarithmic scale).  A value of zero is used whenever no
critical sets were observed.  Of course this is the most conservative
value possible, but there are two other motivations.  First, the trend
of the ``estimated max crit.'' curve starts to fall rapidly at 41 edges,
and a smooth extrapolation would suggest that it plummets, perhaps to
zero, very soon after that point.  Second, when no critical sets were
observed for a given edge size, the probability distribution of the
Bernoulli trial estimation is not ``Gaussian-like'' but is highly
skewed, with a mode (maximum likelihood) of zero critical sets, even
though Eq.~(\ref{eq:lowerconfidence}) may predict a small positive
number.

We emphasize that in the cases where no critical sets were observed,
``estimated max crit.'' merely represents a statistical upper bound
based on the number of random samples we took, meaning it is improbable
that the actual number of critical sets would {\em exceed} that number.
For sizes greater than 41 edges where no critical sets have been
observed, there may be an overriding theoretical reason (that is
currently unknown) that would lead to the actual number of critical sets
being zero.  In that case, ``estimated max crit.'' would get smaller and
smaller, approaching zero, as we increased the number of samples.  But
for any given number of samples, the statistical upper bound is the best
we can do without either a proof that the number of critical sets is
zero or an exhaustive set of samples (which would amount to that proof).
Thus the estimated range on Fig.~\ref{fig:stat} is as objectively
conservative as possible, even though there is subjective evidence,
based on extrapolation at 41 edges, that the actual number of critical
sets becomes identically zero very soon after that point.

\medskip
\noindent{{\bf Acknowledgements} \ {\baselineskip=11pt
One of us (M. P.) would like to thank his host Hossein Sadeghpour
for a support during his stay at ITAMP.
Supported by the {\em US National Science Foundation} through a
grant for the {\em Institute for Theoretical Atomic, Molecular,
and Optical Physics (ITAMP)} at {\em Harvard University and Smithsonian
Astrophysical Observatory} and by the {\em Ministry of Science,
Education, and Sport of Croatia} through {\em Distributed Processing
and Scientific Data Visualization} program and {\em Quantum
Computation: Parallelism and Visualization} project
(082-0982562-3160). Computational support was provided by the
cluster {\em Isabella} of the {\em University Computing Centre} of the
{\em University of Zagreb} and by the {\em Croatian National Grid
Infrastructure}.}

}

\bibliographystyle{elsarticle-num}

\end{document}